\documentclass{ws-ijmpa}
\usepackage[super,compress]{cite}
\usepackage[colorlinks=true, pdfstartview=FitV, linkcolor=red, citecolor=blue, urlcolor=blue]{hyperref}
\usepackage{graphicx}
\usepackage{booktabs}
\begin{document}
\markboth{A. Monnai, B. Schenke and C. Shen}{QCD equation of state at finite chemical potentials for relativistic nuclear collisions}

%
\catchline{}{}{}{}{}
%

\title{QCD EQUATION OF STATE AT FINITE CHEMICAL POTENTIALS FOR RELATIVISTIC NUCLEAR COLLISIONS
}

\author{AKIHIKO MONNAI}

\address{
Department of Mathematical and Physical Sciences, Japan Women's University\\ 
Bunkyo-ku, Tokyo 112-8681, Japan
\\
monnaia@fc.jwu.ac.jp}

\author{BJ\"{O}RN SCHENKE}

\address{Physics Department, Brookhaven National Laboratory\\
Upton, New York 11973, USA\\
bschenke@bnl.gov}

\author{CHUN SHEN}

\address{Department of Physics and Astronomy, Wayne State University\\
Detroit, Michigan 48201, USA\\
RIKEN BNL Research Center, Brookhaven National Laboratory\\
Upton, New York 11973, USA\\
chunshen@wayne.edu}

\maketitle

\begin{history}
\received{\today}
\revised{Day Month Year}
\end{history}

\begin{abstract}
We review the equation of state of QCD matter at finite densities. We discuss the construction of the equation of state with net baryon number, electric charge, and strangeness using the results of lattice QCD simulations and hadron resonance gas models. Its application to the hydrodynamic analyses of relativistic nuclear collisions suggests that the interplay of multiple conserved charges is important in the quantitative understanding of the dense nuclear matter created at lower beam energies. Several different models of the QCD equation of state are discussed for comparison.

\end{abstract}

\ccode{PACS numbers:}


\section{Introduction}	

The collective properties of quantum chromodynamic (QCD) matter have been a topic of great interest in nuclear physics. A milestone has been the discovery of the quark-gluon plasma (QGP) \cite{Letessier:2002gp,Rafelski:2003zz,Yagi:2005yb,Wang:2016opj}, a high-temperature phase of QCD, at the Relativistic Heavy Ion Collider (RHIC) at Brookhaven National Laboratory (BNL) in the year 2000 \cite{Adcox:2004mh,Adams:2005dq,Back:2004je,Arsene:2004fa}. The QGP is speculated to have filled the universe about $10^{-5}$-$10^{-4}$ seconds after the Big Bang. The collider experiments have allowed the quantitative study of QCD matter through comparison of theoretical calculations and experimental data and consequently provided a glimpse of the early universe. The nearly-perfect fluidity and rapid  thermalization of the QGP are major discoveries and have opened up a world of possibilities to study thermodynamics of strongly-interacting elementary particles in collider experiments. 

\begin{figure}[tb]
\centerline{\includegraphics[width=10cm,bb=0 0 942 578]{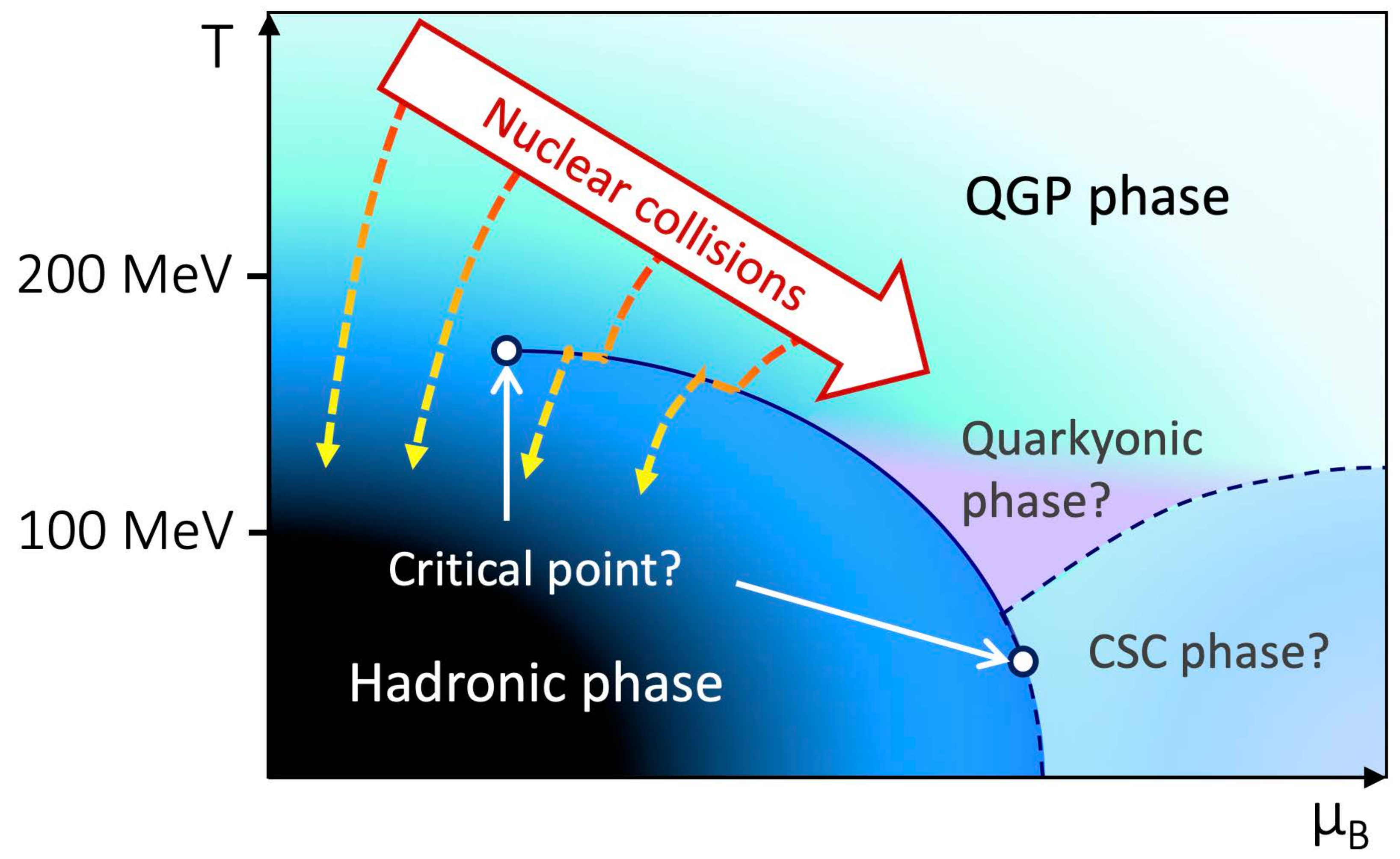}}
\caption{A schematic illustration of the QCD phase diagram and the beam energy scan experiments. The phase structure in the dense regions are conjectured based on model estimations.\label{fig:pd}}
\end{figure}

The high-energy frontier has been explored by the Large Hadron Collider (LHC) at the European Organization for Nuclear Research (CERN), which has been in operation since 2009 \cite{Aamodt:2010pa,ATLAS:2011ah,Chatrchyan:2012wg}. It has extended our experimental knowledge of the QCD phase diagram (Fig.~\ref{fig:pd}) in the direction of temperature, getting closer to the beginning of the universe. The fluidity has been shown to persist at higher temperatures, though the fluid may become less perfect \cite{Gale:2012rq} as the system would be less strongly-coupled. 

The next frontier on the phase diagram is the high-density regime \cite{Fukushima:2010bq, Fukushima:2013rx}, where first principles calculations are known to suffer from the fermion sign problem \cite{deForcrand:2010ys}. Estimations based on the chiral model indicate that the quark-hadron transition turns from a crossover to a first-order phase transition at a finite baryon chemical potential, suggesting the existence of a critical point \cite{Asakawa:1989bq}. Further theoretical model analyses indicate that the QCD phase structure can be quite nontrivial; possible scenarios include the color superconducting (CSC) phase at low temperature and high baryon density, where quarks form a condensate of Cooper pairs \cite{Alford:1997zt,Rapp:1997zu,Alford:1998mk}, the second critical point at the high-density end of the quark-hadron phase boundary implied by the QCD axial anomaly \cite{Hatsuda:2006ps}, the chiral and color superconducting phase transitions enhanced with vector interaction \cite{Kitazawa:2003qmg}, and the quarkyonic phase, suggested by studies in the large $N_c$ limit \cite{McLerran:2007qj}. Exploration of the dense quark matter is of particular importance since the experimental detection of gravitational waves, emerging from e.g. neutron star mergers, now give more stringent constraints on the properties of the compact stars themselves, including the equation of state \cite{TheLIGOScientific:2017qsa,Abbott:2018exr}. See \citen{Baym:2017whm,Dexheimer:2020zzs} for recent reviews. 

\begin{figure}[tb]
\centerline{\includegraphics[height=2.4cm,bb=0 0 503 227]{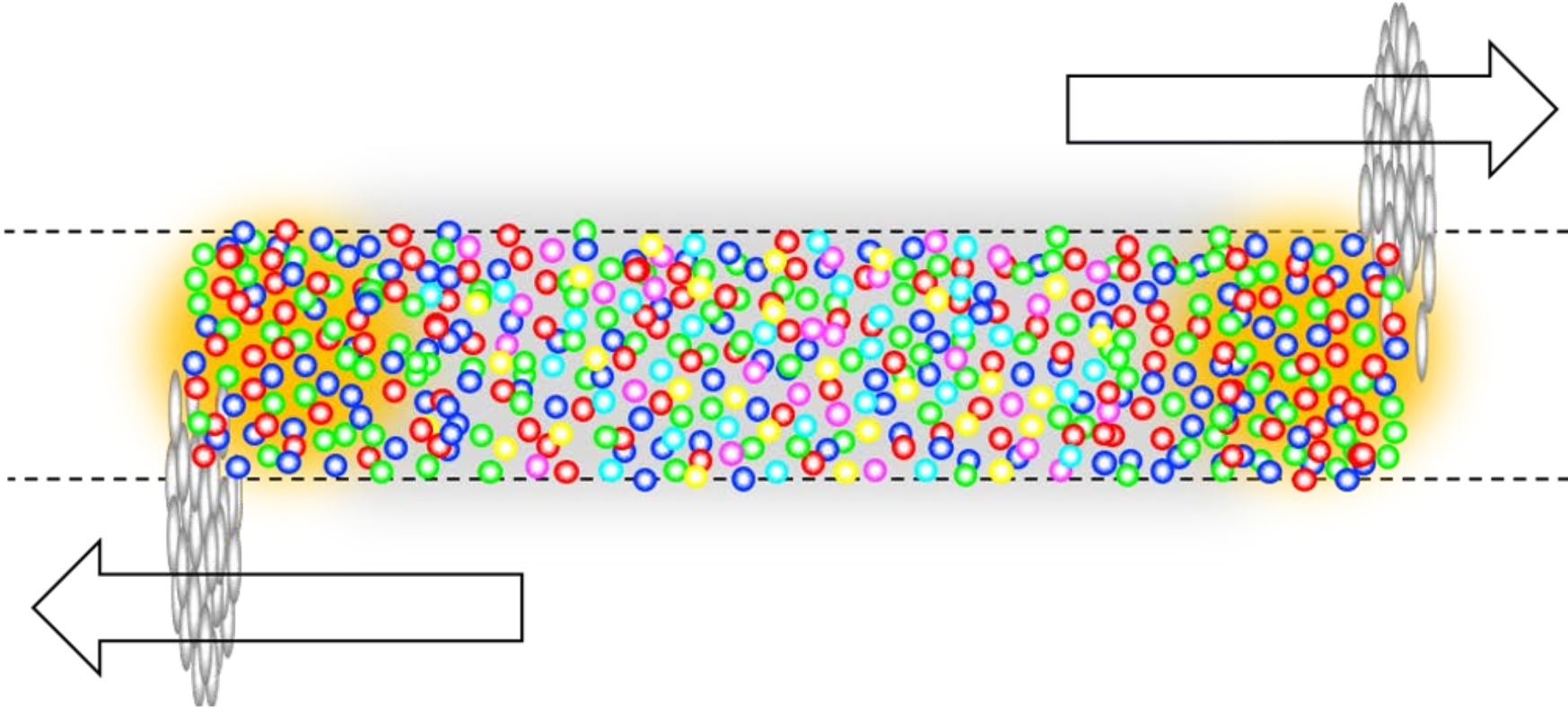}\ \ \ \ \ \
\includegraphics[height=2.4cm,bb=0 0 396 227]{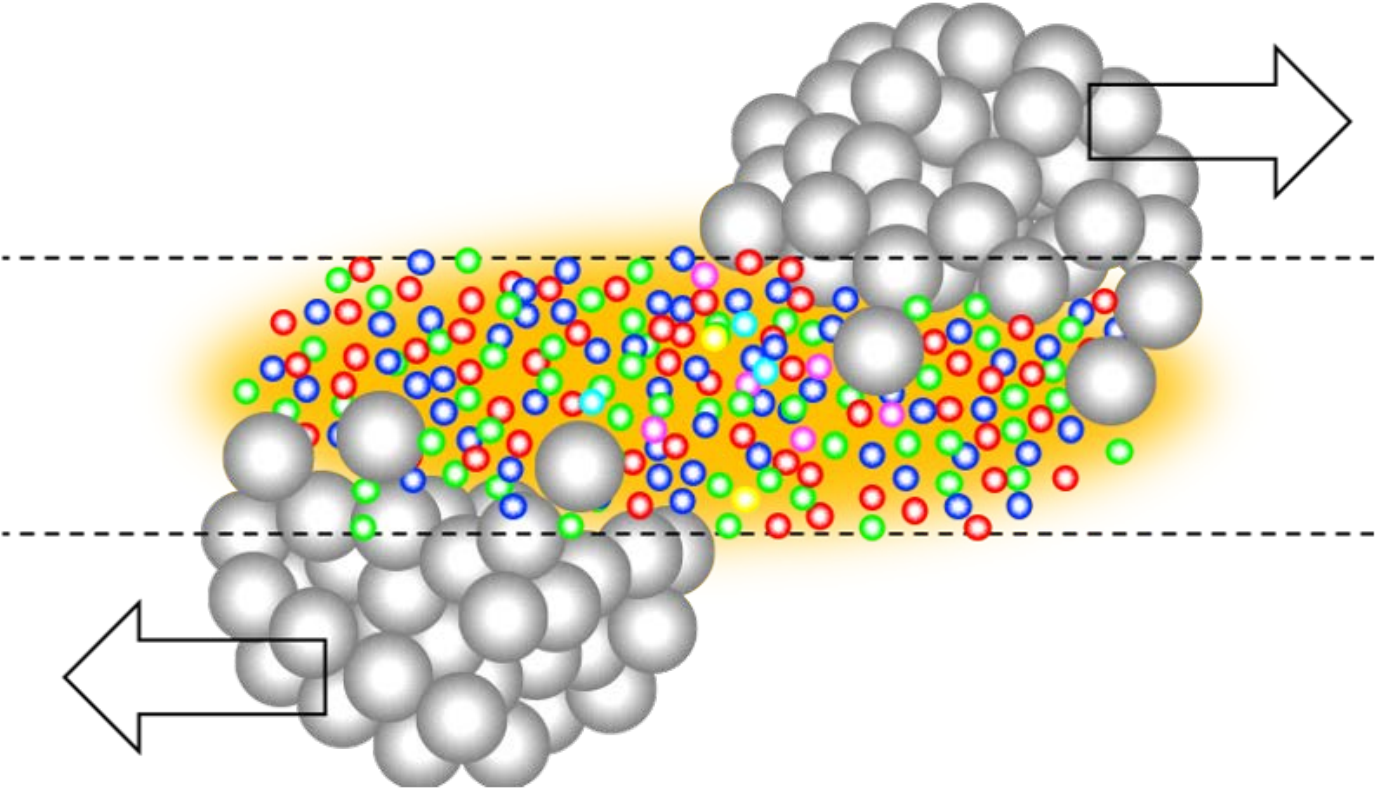}}
\caption{A schematic illustration of the nuclear collisions at high energies where the system has larger temperature and smaller baryon density (left) and at intermediate to low energies where the system has lower temperature and larger baryon density (right).\label{fig:bes}}
\end{figure}

The collider experiments are a powerful tool to obtain bottom-up insight into the QCD matter at finite baryon chemical potential with high precision in controlled environments (Fig.~\ref{fig:bes}). The Beam Energy Scan programs, being preformed at BNL RHIC and planned at various facilities including the GSI Facility for Antiproton and Ion Research (FAIR), JINR Nuclotron-based Ion Collider fAility (NICA), and JAEA/KEK Japan Proton Accelerator Research Complex (J-PARC). The heavy-ion programs at BNL Alternating Gradient Synchrotron (AGS), CERN Super Proton Synchrotron (SPS) and GSI Schwerionensynchrotron 18 (SIS 18), provide complimentary data for understanding the properties of dense quark matter.

One of the most successful models for the description of the dynamical evolution of QGP is the relativistic hydrodynamic model \cite{Kolb:2000fha,Schenke:2010rr}. The observed spectra of hadronic particles up to moderate transverse momenta ($p_T\lesssim 3\,{\rm GeV}$) are known to be in quantitative agreement with hydrodynamic model calculations. Azimuthal momentum anisotropies, characterized with flow harmonics $v_n$ \cite{Ollitrault:1992bk, Poskanzer:1998yz,Takahashi:2009na,Alver:2010gr}, are considered to be one of the most prominent pieces of evidence for the nearly-perfect fluidity of the produced QCD medium, because they are found to clearly reflect the geometrical anisotropy of the overlap region of colliding nuclei, implying that the system is strongly coupled. The physics of QCD enters the model through the equation of state -- and the transport coefficients in off-equilibrium cases -- along with details of initial conditions. Thus, once a realistic initial geometry is given, it is in principle possible to extract information on the QCD equation of state by comparing numerical results based on trial input with experimental data \cite{Pratt:2015zsa,Sangaline:2015isa,Bernhard:2016tnd,Pang:2016vdc,Monnai:2017cbv,Paquet:2017mny}.

The equation of state is a fundamental relation among thermodynamic variables. Earliest studies of the QCD equation of state date back to the MIT bag model \cite{Chodos:1974je,Chodos:1974pn}, where confinement is introduced phenomenologically. Hadrons are treated as quarks in bags within the QCD vacuum. Consequently, the model has a first-order phase transition between the hadron and QGP phases. Since then, our understanding of  QCD thermodynamics has been deepened with the advent of model approaches such as the potential model \cite{DeRujula:1975qlm} and the Nambu-Jona-Lasinio model \cite{Nambu:1961tp,Nambu:1961fr}. A breakthrough was brought when first principle calculations became possible with the advancement of the computational method of lattice QCD (at zero chemical potentials). SU(3) pure glue studies predict a first-order QCD phase transition while more realistic (2+1)-flavor calculations suggest a crossover transition, implying the importance of quark contributions in the phenomenon \cite{Brown:1990ev,AliKhan:2000wou,Aoki:2006we}. 
State-of-the-art lattice QCD simulations with a physical pion mass provide high-precision results of the QCD equation of state over a wide-range of temperatures \cite{Borsanyi:2013bia, Bazavov:2014pvz,Ding:2019prx}. The lattice QCD equation of state, when embedded in a hydrodynamic model, is known to reproduce the experimental data of nuclear collisions at top RHIC and LHC energies well.

It was considered in the earlier days of QGP phenomenology that the fluidity appears only at and above energies around $\sqrt{s_{NN}} = \mathcal{O}(10^2)$ GeV. The net baryon density in most cases was not considered important because it would be small and have negligible effects for such systems, except at forward rapidities. As the hydrodynamic model became more sophisticated, on the other hand, it was rediscovered that the hydrodynamic description can be valid for hadronic yields at lower energies, down to $\sqrt{s_{NN}} = \mathcal{O}(10)$ GeV, which is the typical energy scale covered by the beam energy scan programs\cite{Shen:2020gef, Shen:2020mgh}. This may be partially owing to the fact that one has a better understanding of non-equilibrium processes -- initial dynamics of local equilibration, viscosity and diffusion, and hadronic transport, which occur before, during, and after the hydrodynamic evolution, respectively -- and can now show that hydrodynamics, which is based on the idea of local equilibrium, is compatible with experimental results.

For a quantitative description of nuclear collisions in the beam energy scan programs, one needs the equation of state at finite densities \cite{Nonaka:2004pg,Bluhm:2004xn,Bluhm:2007nu,Dexheimer:2009hi,Steinheimer:2010ib,Huovinen:2011xc,Hempel:2013tfa,Albright:2014gva,Albright:2015uua,Rougemont:2017tlu,Critelli:2017oub,Vovchenko:2017gkg,Parotto:2018pwx,Vovchenko:2018zgt,Fu:2018qsk,Fu:2018swz,Motornenko:2018hjw,Plumberg:2018fxo} as input for hydrodynamic simulations. First principle calculations are known to be challenging at finite densities, owing to the aforementioned sign problem. Several intriguing methods to circumvent the sign problem have been proposed, such as the Taylor expansion method \cite{Gavai:2001fr,Allton:2002zi}, the reweighting method \cite{Fodor:2004nz,Allton:2002zi,Fodor:2001au}, the imaginary chemical potential method \cite{deForcrand:2002hgr,DElia:2002tig,Gunther:2016vcp}, the complex Langevin method \cite{Parisi:1984cs,Klauder:1985,Ambjorn:1985iw,Nagata:2016vkn}, the Lefschetz thimble method \cite{Pham:1983,Witten:2010cx}, and the path optimization method \cite{Mori:2017pne,Mori:2017nwj}, but so far no complete description is available at small temperatures and large chemical potentials.

In this review, we will discuss the phenomenological construction of the QCD equation of state at finite chemical potentials for relativistic nuclear collisions. Oftentimes, only net baryon number is taken into account as the conserved charge in the equation of state, especially when hydrodynamic modeling is concerned. We review the \textsc{neos} model \cite{Monnai:2019hkn,Monnai:2020pcw} based on the lattice QCD equation of state and susceptibilities at vanishing densities from Refs. \citen{Bazavov:2014pvz,Bazavov:2012jq, Ding:2015fca, Bazavov:2017dus, Sharma} (see also Refs. \citen{Borsanyi:2011sw,Bellwied:2015lba,Borsanyi:2018grb, Ding:2019prx, Borsanyi:2020fev}), and the hadron resonance gas equation of state, that include three conserved charges relevant in nuclear collisions: net baryon (B), electric charge (Q), and strangeness (S) as a successor to the version including only net baryon chemical potentials \cite{Denicol:2015nhu,Monnai:2015sca,Shen:2017ruz,Shen:2017bsr,Denicol:2018wdp,Shen:2018pty,Gale:2018vuh}. The selection of conserved charges is based on the assumption that only light quarks ($u$, $d$, and $s$) would thermalize in nuclear collisions. It has been employed in recent hydrodynamic model analyses \cite{Shen:2020jwv,Oliinychenko:2020znl,Zhao:2020irc}. A similar approach also has recently been proposed in Ref.~\citen{Noronha-Hostler:2019ayj}, and the importance of multiple conserved charges has been discussed in various situations before our model realization \cite{Fukushima:2009dx,Toublan:2004ks,Xu:2011pz,Ohnishi:2011jv,Kamikado:2012bt,Ueda:2013sia,Barducci:2004tt,Nishida:2003fb,Son:2000xc,Karpenko:2015xea,Hatta:2015era,Hatta:2015hca,Andronic:2005yp,Bazavov:2012vg}. We demonstrate by explicit calculations within a hydrodynamic model that the description of experimental data is improved by our comprehensive treatment of the conserved charges in the construction of the equation of state. Finally, we compare different models for the QCD equation of state, both at zero and finite densities, and present conclusions and summary.

The natural units $c = \hbar = k_B = 1$ and the mostly-minus Minkowski metric $g^{\mu \nu} = \mathrm{diag}(+,-,-,-)$ are used.

\section{Status}

We review the status of the study of the QCD equation of state. QCD thermodynamics is a topic of interest to a broad range of studies from nuclear physics to particle physics to astrophysics. Here we focus on the phenomenological equations of state intended for use in hydrodynamic studies of relativistic nuclear collisions.

\subsection{From bag model to lattice QCD}

Early hydrodynamic models often employed the equation of state inspired by the MIT bag model, such as EOS Q, which has a first-order phase transition at zero densities. The hadronic phase is described using a resonance gas and the QGP phase using a parton gas with a bag constant \cite{Sollfrank:1996hd,Nonaka:2000ek,Teaney:2000cw,Kolb:2000fha}. 
A finite net baryon density was relatively easy to implement in such models, though it was neglected in many cases, because it would have small effects around mid-rapidity at top RHIC energies. These equations of state are, despite involving a first-order phase transition, able to reproduce the experimental data of hadronic spectra and elliptic flow reasonably well with an appropriate choice of initial conditions in inviscid models. 

Crossover-like equations of state have also been discussed in the literature. Early studies include a functional parametrization of pioneering lattice QCD results \cite{Redlich:1985uw,Blaizot:1987nd,Rischke:1995cm} by matching a parton gas with a pion gas equation of state
and encoding the details of the transition using the choice of the connection width $\Delta T$ (defined later as in 
Eq.\,\eqref{eq:cfunction}). 
A more sophisticated equation of state was developed by connecting the results of the hadron resonance gas and an effective theory for finite temperature SU(3) gauge theory, \cite{Kajantie:1997tt, Laine:2006cp}
and was used for viscous hydrodynamic analyses \cite{Romatschke:2007mq}. 
With the advent of first principle calculations, the connection of the results of lattice QCD simulations and hadron resonance gas became a topic of interest \cite{Biro:2006sv,Chojnacki:2007jc,Chojnacki:2007rq,Huovinen:2009yb,Bluhm:2013yga,Alba:2017hhe,Auvinen:2020mpc}. A variation of such approach includes the quasi-particle model fit to the lattice QCD data \cite{Bluhm:2004xn, Huovinen:2005gy}.
Sometimes the lattice QCD equation of state is used directly at vanishing densities down into the hadronic phase, though caution is needed because the energy-momentum conservation at particlization is no longer automatically guaranteed if the hadron resonance gas description is not used \cite{Cooper:1974mv}, and the inconsistency may be hidden by the normalization of initial conditions. Additionally, the lattice QCD data in the continuum limit typically have uncertainty bands of a few percent, though they have been improved considerably in recent simulations. 

As mentioned earlier, the lattice-based QCD equation of state is considered to give an accurate description of the hot matter created and observed in the collider experiments at RHIC and LHC, where the conserved charges can be neglected \cite{Pratt:2015zsa,Sangaline:2015isa,Bernhard:2016tnd,Pang:2016vdc,Monnai:2017cbv,Paquet:2017mny}. It is important to next elucidate the high density regions of the QCD phase diagram for fully utilizing the data from the ongoing and upcoming beam energy scan programs and for understanding microscopic properties of the QCD matter near equilibrium.

\subsection{Equation of state at finite densities}
The finite-density version of a hybrid equation of state s95p-v1 \cite{Huovinen:2011xc} is one of the pioneering studies to use the coefficients of the Taylor expansion method for construction. A temperature shift was introduced to the susceptibilities estimated in lattice QCD calculations with larger than physical pion mass, which tend to produce a higher $T_c$ than those with physical pion mass, for smooth matching to the resonance gas results. As the lattice QCD calculations improved, one has begun to use the bare result of the baryon susceptibility in hydrodynamic simulations \cite{Monnai:2012jc,Monnai:2014qaa}. The connection of the lattice QCD results using the physical pion mass to the hadron resonance gas results has been discussed at finite density of net baryons \cite{Monnai:2015sca} and of net baryons, electric charge and strangeness \cite{Monnai:2019hkn,Noronha-Hostler:2019ayj}. 

The implementation of a critical point in the equation of state is also a topic of importance \cite{Asakawa:1989bq,Halasz:1998qr,Stephanov:2004wx}.
The 3-dimensional Ising model is often used for this purpose because it is considered to be in the same universality class as QCD \cite{Nonaka:2004pg,Parotto:2018pwx}.
The rescaled magnetic field and the reduced temperature in the latter model \cite{Guida:1996ep} are mapped onto the reduced temperature and (baryon) chemical potential in QCD, respectively.
Experimental elucidation of the critical point is a long-standing goal to which no complete answer is available yet \cite{Adamczyk:2011aa,Adamczyk:2013dal}.

Perturbative QCD calculations have also been improved to include higher order contributions \cite{Shuryak:1977ut, Kapusta:1979fh, Toimela:1982hv, Arnold:1994ps, Arnold:1994eb, Zhai:1995ac, Kajantie:2002wa}, though the convergence of the weak-coupling expansion for the pressure is slow, even at the $g^6$ order, and the dependence on the renormalization scale is large \cite{Su:2012iy}. In light of this situation, improved versions of the perturbation theory have been proposed, such as the two-particle irreducible (2PI) formalism \cite{Blaizot:2003tw,Kraemmer:2003gd,Andersen:2004fp} and hard thermal loop (HTL) perturbation theory \cite{Andersen:1999fw,Andersen:1999sf,Andersen:1999va}. There are quantitative studies to match perturbative results \cite{Vuorinen:2002ue,Vuorinen:2003fs,Haque:2012my,Haque:2013sja,Andersen:2015eoa} to the hadron resonance gas ones with the help of lattice QCD data to approach the finite density regime \cite{Albright:2014gva}. 

The effective model approaches to the finite-density phase structure include the Nambu-Jona-Lasinio model with the Polyakov loop \cite{Fukushima:2003fw,Fukushima:2009dx,Ratti:2005jh,Steinheimer:2010ib}, polyakov loop enhanced quark-meson models \cite{Fu:2018qsk}, and a more phenomenological quasi-particle model, \cite{Bluhm:2007nu} where the result of Ref.~\citen{Steinheimer:2010ib} has been employed in one of the first modern hydrodynamic simulations of the beam energy scan experiments \cite{Karpenko:2015xea}. 

An alternative approach to describe strongly-coupled matter is via holographic gauge-string duality \cite{Maldacena:1997re,Gubser:1998bc,Witten:1998qj}. The original anti-de Sitter/conformal field theory (AdS/CFT) correspondence is conjectured for the $\mathcal{N}=4$ super Yang-Mills theory, which is scale invariant and thus has no phase transition. The primary role of the conjecture in the phenomenology of nuclear collisions is perhaps the prediction of transport coefficients \cite{Kovtun:2004de,Buchel:2007mf,Natsuume:2007ty,Romatschke:2009kr}, for which first principle calculations are difficult. Extensions of this method to non-conformal theories have been proposed in order to preserve consistency with the thermodynamic properties of QCD. Such examples include the Einstein-Maxwell-Dilation model \cite{Rougemont:2015wca,DeWolfe:2010he,DeWolfe:2011ts} which can reproduce the known lattice QCD data.

The QCD equation of state is also a topic of importance for compact stars. The typical chemical potential is larger and the temperature is smaller in such systems compared with those in nuclear collisions \cite{Nara:1997zp}, though it may be possible to have occasional baryonic dense spots in the latter through event-by-event fluctuations. There have been extensive studies on the neutron star equation of state -- see, \textit{e.g.}, Refs.~\citen{Baym:2017whm,Dexheimer:2020zzs} for recent reviews. In addition to the intriguing observation regarding the Shapiro delay\cite{Demorest:2010bx}, the experimental discovery of gravity waves has brought a plethora of new data, which can constrain the nuclear equation of state in the cold and dense regime \cite{Abbott:2018exr}.

\section{NEOS -- hybrid QCD equation of state}	

We discuss the construction of the QCD equation of state at finite chemical potentials of net baryon number, electric charge, and strangeness. \textsc{neos} is an equation of state model, which takes one of the latest lattice QCD equations of state at vanishing chemical potentials as a baseline. Following the Taylor expansion method, the second- and fourth-order diagonal and off-diagonal susceptibilities are implemented. This expansion method has the advantage of being able to express the thermodynamic variables at finite density with those at zero density. On the other hand, one needs an additional prescription at low temperatures, because the Taylor expansion becomes less reliable when the chemical potential over temperature ratio is large \cite{Karsch:2010hm}. Thus, the hadron resonance gas model, which is a framework to understand the low-temperature QCD system in terms of stable hadrons and meta-stable resonances, is used at lower temperatures, and its pressure is matched to the lattice-based pressure near the crossover.

There are additional motivations for the connection procedure. First, all the thermodynamic variables and second- and fourth- order susceptibilities of the hadron resonance gas model are known to show excellent agreement with those of lattice QCD. The fact that the hadron resonance gas model shares basic thermodynamic properties with the first principle calculation motivates one to assume that the model captures essential physics. Second, the success of hydrodynamic modeling relies on the hadron resonance gas picture when the flow field is converted into hadronic particles.\footnote{It should be noted that the experimental data for particle spectra, chemical ratios, and chemical freeze-out indicate that the concept of temperature is valid near particlization in nuclear collisions, even though there are arguments regarding hydrodynamization without thermalization at the earliest stage of hydrodynamic evolution \cite{Romatschke:2016hle,Heller:2016rtz,Florkowski:2017olj}. In anisotropic hydrodynamics thermodynamic variables can receive modifications from local momentum anisotropies \cite{Florkowski:2010cf,Martinez:2010sc,Bazow:2013ifa,Alqahtani:2018fqi,Alqahtani:2020paa}.} The Cooper-Frye prescription \cite{Cooper:1974mv} for particlization also requires that the equation of state in the hydrodynamic evolution is the same as that in the hadronic transport model to allow for energy-momentum and charge conservation. Direct use of the hadronic resonance gas equation of state at low temperatures is the  most practical way to achieve this.

\subsection{Lattice QCD equation of state in the Taylor expansion method}

The higher temperature side of the equation of state is constructed using the Taylor-expanded pressure $P_\mathrm{lat}$, 
\begin{eqnarray}
\frac{P_\mathrm{lat}}{T^4} &=& \frac{P_0}{T^4} + \sum_{l,m,n} \frac{\chi^{B,Q,S}_{l,m,n}}{l!m!n!} \bigg( \frac{\mu_B}{T} \bigg)^{l}  \bigg( \frac{\mu_Q}{T} \bigg)^{m}  \bigg( \frac{\mu_S}{T} \bigg)^{n}, 
\label{Psus}
\end{eqnarray}
where $P_0$ and $\chi^{B,Q,S}_{l,m,n}$ are the pressure and $(l+m+n)$-th order susceptibilities at zero chemical potentials, calculated in lattice QCD simulations. $T$ is the temperature and $\mu_B$, $\mu_Q$, $\mu_S$ are the chemical potentials for net baryon, electric charge, and strangeness, respectively. They are related to the quark chemical potentials as 
\begin{eqnarray}
\mu_u &=& \frac{1}{3}\mu_B + \frac{2}{3} \mu_Q, \label{eq:muu} \\
\mu_d &=& \frac{1}{3}\mu_B - \frac{1}{3} \mu_Q, \label{eq:mud} \\
\mu_s &=& \frac{1}{3}\mu_B - \frac{1}{3} \mu_Q - \mu_S. \label{eq:mus}
\end{eqnarray}
The susceptibilities can be expressed as
\begin{equation}
  \chi_{l,m,n}^{B,Q,S} = \frac{\partial^l \partial^m \partial^n P(T,\mu_B,\mu_Q,\mu_S)/T^4}{\partial(\mu_B/T)^l\partial(\mu_Q/T)^m\partial(\mu_S/T)^n}\bigg|_{\mu_{B,Q,S}=0}.
\end{equation}
$l+m+n$ is constrained by the matter-antimatter symmetry to be even. One can alternatively consider isospin instead of electric charge.

\subsection{Hadron resonance gas equation of state}

The hadron resonance gas picture is used for calculating the lower temperature side of the equation of state. Its pressure reads
\begin{eqnarray}
P_\mathrm{had} &=& \pm T \sum_i \int \frac{g_i d^3p}{(2\pi)^3} \ln [1 \pm e^{-(E_i-\mu_i)/T} ]\nonumber\\
&=& \sum_i \sum_k (\mp1)^{k+1} \frac{1}{k^2} \frac{g_i}{2\pi^2} m_i^2 T^2 e^{k\mu_i/T} K_2\bigg(\frac{k m_i}{T}\bigg), \label{eq:P_had}
\end{eqnarray}
where $E_i = \sqrt{p^2+m_i^2}$ is the energy, $m_i$ is the mass, $g_i$ is the degeneracy, and $\mu_i$ is the chemical potential of the $i$-th hadronic species. $\mu_i$ can be expressed as $\mu_i = B_i \mu_B + Q_i \mu_Q + S_i \mu_S$ using the quantum numbers $B_i$, $Q_i$, and $S_i$ for net baryon, electric charge, and strangeness. $K_2(x)$ is the modified Bessel function of the second kind. The expansion with $k$ takes account of the correction of quantum statistics. It is usually sufficient to consider the $k \leq 3$ terms for pions, the $k \leq 2$ terms for kaons, and the $k = 1$ term for heavier particles. The Boltzmann limit corresponds to the $k=1$ case. 
We treat hadrons as on-shell particles and do not include spectral functions for the resonance states in our model \cite{Tanabashi:2018oca}.

\subsection{Hybrid equation of state} 

The \textsc{neos} equation of state is obtained by connecting the ones from lattice QCD and the hadron resonance gas model. The pressure is given as
\begin{eqnarray}
\frac{P}{T^4} &=& \frac{1}{2}[1- f(T,\mu_B,\mu_Q,\mu_S)] \frac{P_{\mathrm{had}}(T,\mu_B,\mu_Q,\mu_S)}{T^4} \nonumber \\
&+&\frac{1}{2}[1+ f(T,\mu_B,\mu_Q,\mu_S)] \frac{P_{\mathrm{lat}}(T,\mu_B,\mu_Q,\mu_S)}{T^4} , \label{eq:econ}
 \end{eqnarray}
 where the connecting function $f$ should satisfy $f\to 1$ and $f\to -1$ in the high and low temperature limits, respectively. Here we choose a smooth hyperbolic function
\begin{equation}
f(T,\mu_B,\mu_Q,\mu_S) = \tanh \bigg[ \frac{T-T_c(\mu_B)}{\Delta T_c} \bigg].
\label{eq:cfunction}
\end{equation} 
$T_c(\mu_B)$ is the connecting temperature for which we use $T_c(\mu_B) = 0.16\ \mathrm{GeV} - 0.4 \,(0.139\ \mathrm{GeV}^{-1} \mu_B^2 + 0.053\ \mathrm{GeV}^{-3} \mu_B^4)$ motivated by the $\mu_B$ dependence of the chemical freeze-out line \cite{Cleymans:2005xv}. The connecting width is chosen to be $\Delta T_c = 0.1 T_c (0)$. The dependencies on electric charge and strangeness chemical potentials are assumed to be small and neglected here. The choices of possible parameter values and their effects are limited for the following reasons. First, the thermodynamic conditions
\begin{eqnarray}
\frac{\partial^2 P}{\partial T^2} &=& \frac{\partial s}{\partial T} > 0, \label{eq:tcc1} \\ 
\frac{\partial^2 P}{\partial \mu_{B,Q,S}^2} &=& \frac{\partial n_{B,Q,S}}{\partial \mu_{B,Q,S}} > 0 . \label{eq:tcc2}
\end{eqnarray}
have to be imposed near the connection range because they would no longer be trivially satisfied when two different frameworks are being connected by another function. The procedure leaves a narrow window for the possible choice of parameters. Second, the fact that the lattice QCD and hadron resonance gas equations of state match over a finite temperature range implies that the overall thermodynamic properties of the system do not and should not depend on the detailed parameter choice. 

The above procedure gives a crossover equation of state by construction. One may argue that there could be a critical point in the accessible range of the QCD phase diagram. We consider the crossover-type equation of state here to allow for baseline hydrodynamic calculations without critical behavior. Future experimental observation of deviations from that baseline can then be analyzed to deduce the existence and location of the QCD critical point. If one introduces $f$ with a non-differentiable kink, an equation of state with a first-order phase transition is easily obtained \cite{Plumberg:2018fxo}.

It is useful here to introduce basic thermodynamic relations for estimating other macroscopic variables. The entropy density $s$, the net baryon, electric charge, and strangeness densities $n_{B,Q,S}$, the energy density $e$, and the sound velocity $c_s$ are obtained via
\begin{eqnarray}
s &=& \left. \frac{\partial P}{\partial T} \right|_{\mu_{B},\mu_{Q},\mu_{S}}, \ \
n_{B} = \left. \frac{\partial P}{\partial \mu_{B}} \right|_{T,\mu_{Q},\mu_{S}}, \\
n_{Q} &=& \left. \frac{\partial P}{\partial \mu_{Q}} \right|_{T,\mu_{B},\mu_{S}}, \ \
n_{S} = \left. \frac{\partial P}{\partial \mu_{S}} \right|_{T,\mu_{B},\mu_{Q}}, \\
e &=& Ts - P + \mu_B n_B + \mu_Q n_Q+ \mu_S n_S, \\
c_s^2 &=& \left. \frac{\partial P}{\partial e} \right|_{n_B,n_Q,n_S} + \frac{n_B}{e+P} \left. \frac{\partial P}{\partial n_B} \right|_{e, n_Q,n_S} \nonumber \\
&+& \frac{n_Q}{e+P} \left. \frac{\partial P}{\partial n_Q} \right|_{e, n_B,n_S}+ \frac{n_S}{e+P} \left. \frac{\partial P}{\partial n_S} \right|_{e, n_B,n_Q}, \label{eq:cs2}
\end{eqnarray}
for the system with multiple conserved charges.

\subsection{Multiple charges in nuclear collisions}

The strangeness density in nuclear collisions on average is vanishing because the colliding nuclei are net strangeness free. This is called the strangeness neutrality condition. The condition leads to positive strangeness chemical potential in the presence of positive baryon chemical potential, because the number of strange quarks would exceed that of anti-quarks in the QGP phase if $\mu_S = 0$ was assumed. An interpretation based on the parton picture is that $\mu_S \sim \mu_B/3$ follows from Eq.~(\ref{eq:mus}) when $\mu_Q \sim 0$. For the hadronic phase, the strangeness chemical potential can be suppressed because the lightest baryon with strangeness is $\Lambda$, the mass of which is already large compared with the temperature of the system.

The electric charge density is related to the net baryon density via the proton-to-nucleon number ratio $Z/A$. $Z/A$ of the nuclei used or planned in the collider experiments at RHIC and LHC are listed in Table~\ref{table:za}. The primarily-used heavy ions Au and Pb have $Z/A \approx 0.4$. For neutron-rich nuclei, the chemical potential of $d$ quarks is larger than that of $u$ quarks, \textit{i.e.}, $\mu_d = \mu_B/3 - \mu_Q/3 > \mu_u = \mu_B/3 + 2\mu_Q/3$, which implies that $\mu_Q < 0$ when $\mu_B > 0$ in the QGP phase. The trend remains in the hadronic phase because negative pions would be abundant compared with positive pions, which leads to $\mu_{\pi^-} = - \mu_Q > \mu_{\pi^+} = \mu_Q$. One would have the opposite situation $\mu_Q > 0$ for proton-rich nuclei, which are relevant in smaller systems.

\begin{table}[tb]
\tbl{Number ratios of protons to nucleons $Z/A$ for the nuclei used or planned at RHIC and LHC.} 
{\begin{tabular}{ccccccc}
\toprule
Nucleus &  $^{1}_{1}$H &$^{2}_{1}$H & $^{3}_{2}$He & $^{8}_{16}$O & $^{27}_{13}$Al & $^{63}_{29}$Cu  \\ \midrule
$Z/A$  & 1.000 & 0.500& 0.667& 0.500& 0.481& 0.460 \\ \midrule
Nucleus &  $^{96}_{40}$Zr & $^{96}_{44}$Ru &$^{127}_{54}$Xe &$^{197}_{\ 79}$Au & $^{208}_{\ 82}$Pb & $^{238}_{\ 92}$U \\ \midrule
$Z/A$  & 0.417 & 0.458  & 0.425  & 0.401 & 0.394 & 0.387 \\ \botrule
\end{tabular}
\label{table:za}}
\end{table}

\subsection{Numerical construction}

Results of (2+1)-flavor lattice QCD simulations are used to evaluate the pressure \cite{Bazavov:2014pvz} and the second- and fourth-order susceptibilities \cite{Bazavov:2012jq, Ding:2015fca, Bazavov:2017dus, Sharma} at vanishing densities in the numerical construction of the hybrid equation of state. In addition, $\chi_{6}^B$, $\chi_{5,1}^{B,Q}$, and $\chi_{5,1}^{B,S}$ of the sixth-order susceptibilities are phenomenologically introduced for a proper matching of the thermodynamic variables because the results of the Taylor expansion method of lattice QCD simulations cannot be na\"{i}vely used when they have large error bars, as small displacement of the crossover temperature can lead to unphysical gaps in thermodynamic quantities when  $\mu_B/T$ is large. The Stefan-Boltzmann limits are used as anchors on the high temperature side so the basic thermodynamic properties are preserved when lattice QCD data points are scarce. Those treatments could be improved in the future when more data become available. The functional forms for the parametrization of all the susceptibilities used in the model are found in Ref.~\citen{Monnai:2019hkn}. 

The hadron resonance gas model includes all the hadrons and resonances which have $u$, $d$ and/or $s$ as constituent components and have masses smaller than 2 GeV in the Particle Data Group list \cite{Tanabashi:2018oca}. The pressure and susceptibilities up to the fourth order are found to agree well with those of lattice QCD calculations. 

The following three situations are simulated: (i) the conventional situation $\mu_S = \mu_Q = 0$ where only the net baryon number is considered as conserved charge, (ii) the situation with the strangeness neutrality condition $n_S = 0$ and vanishing electric charge chemical potential $\mu_Q = 0$, and (iii) the realistic situation in collisions of heavy nuclei where $n_S = 0$ and $n_Q = 0.4\,n_B$. They are labeled as \textsc{neos} B, \textsc{neos} BS, and \textsc{neos} BQS, respectively, in the article.  

\begin{figure}[tb]
\centerline{\includegraphics[width=2.4in,bb=0 0 360 252]{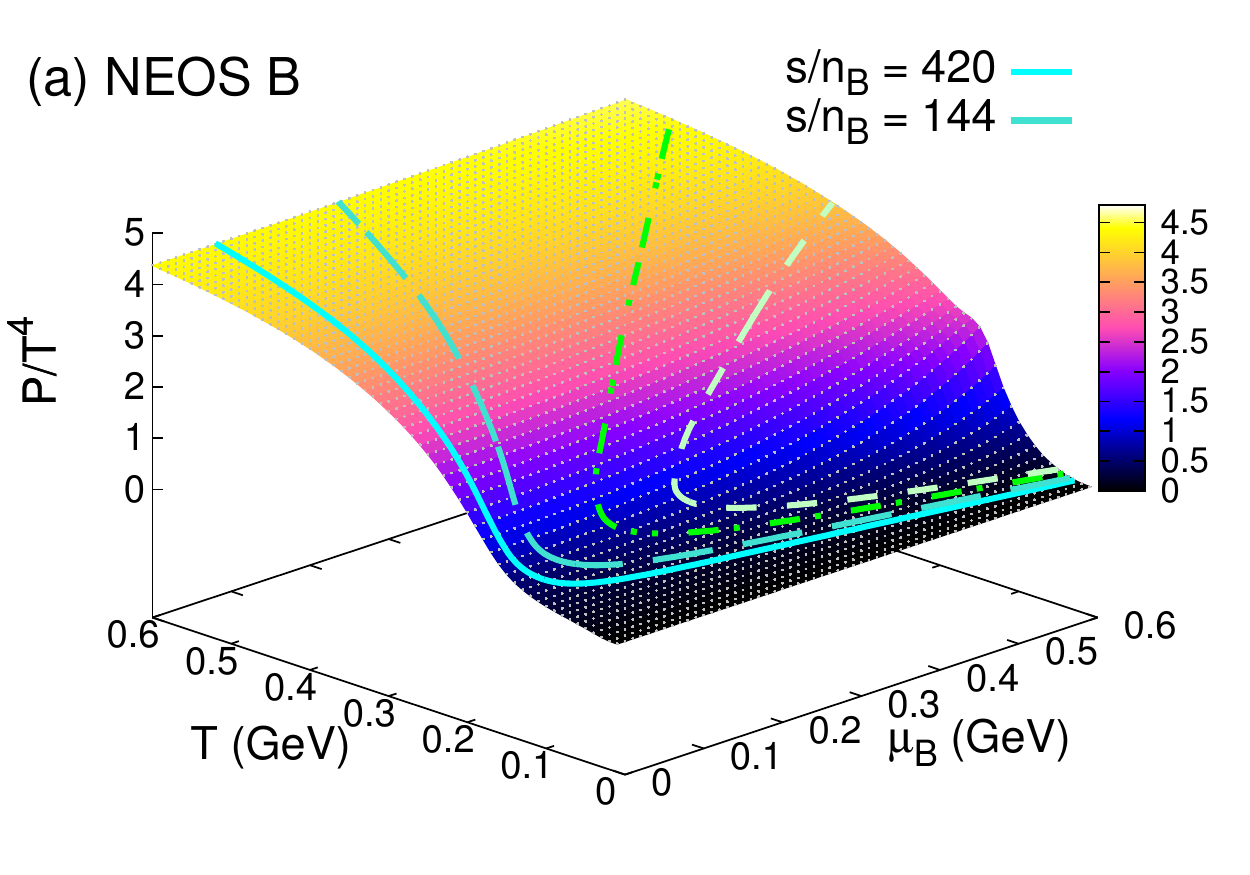}
\includegraphics[width=2.4in,bb=0 0 360 252]{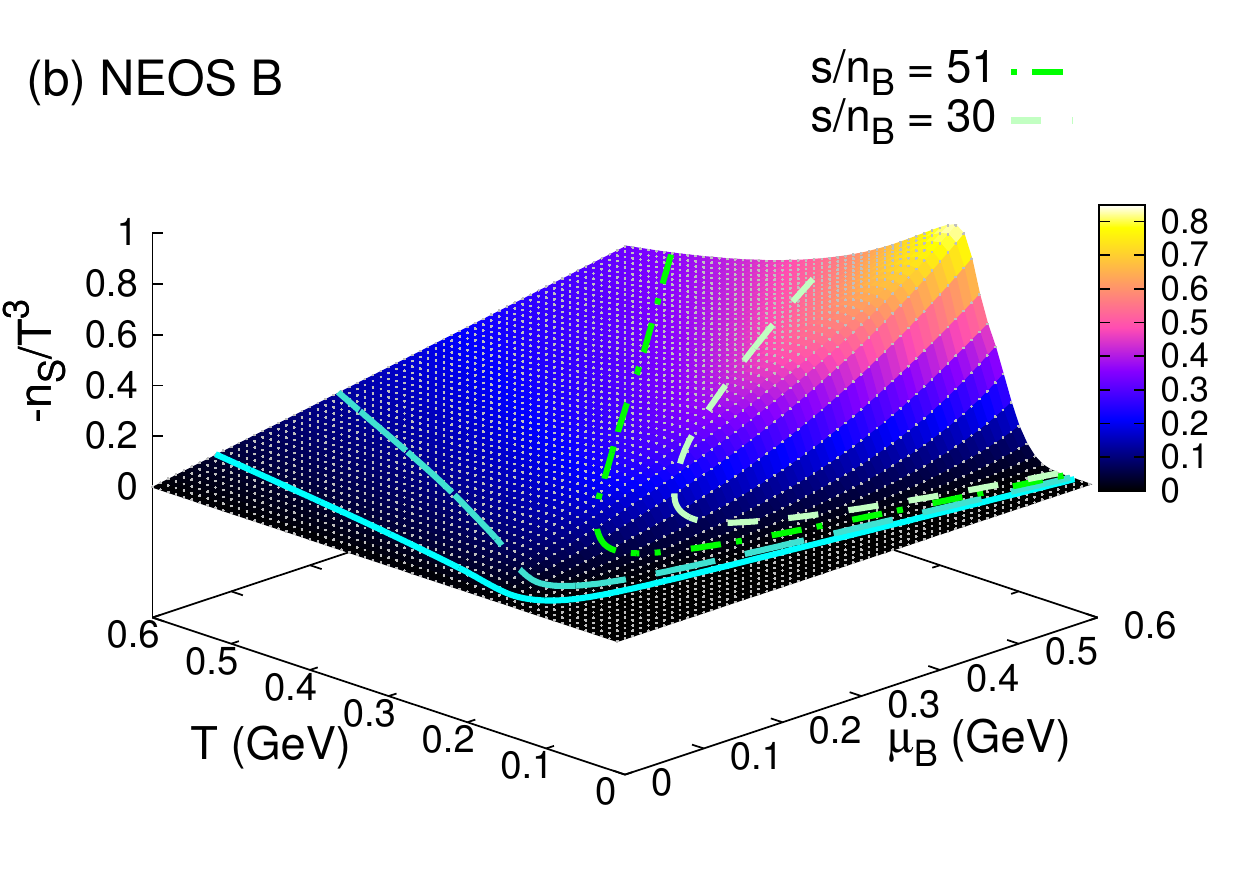}}
\centerline{\includegraphics[width=2.4in,bb=0 0 360 252]{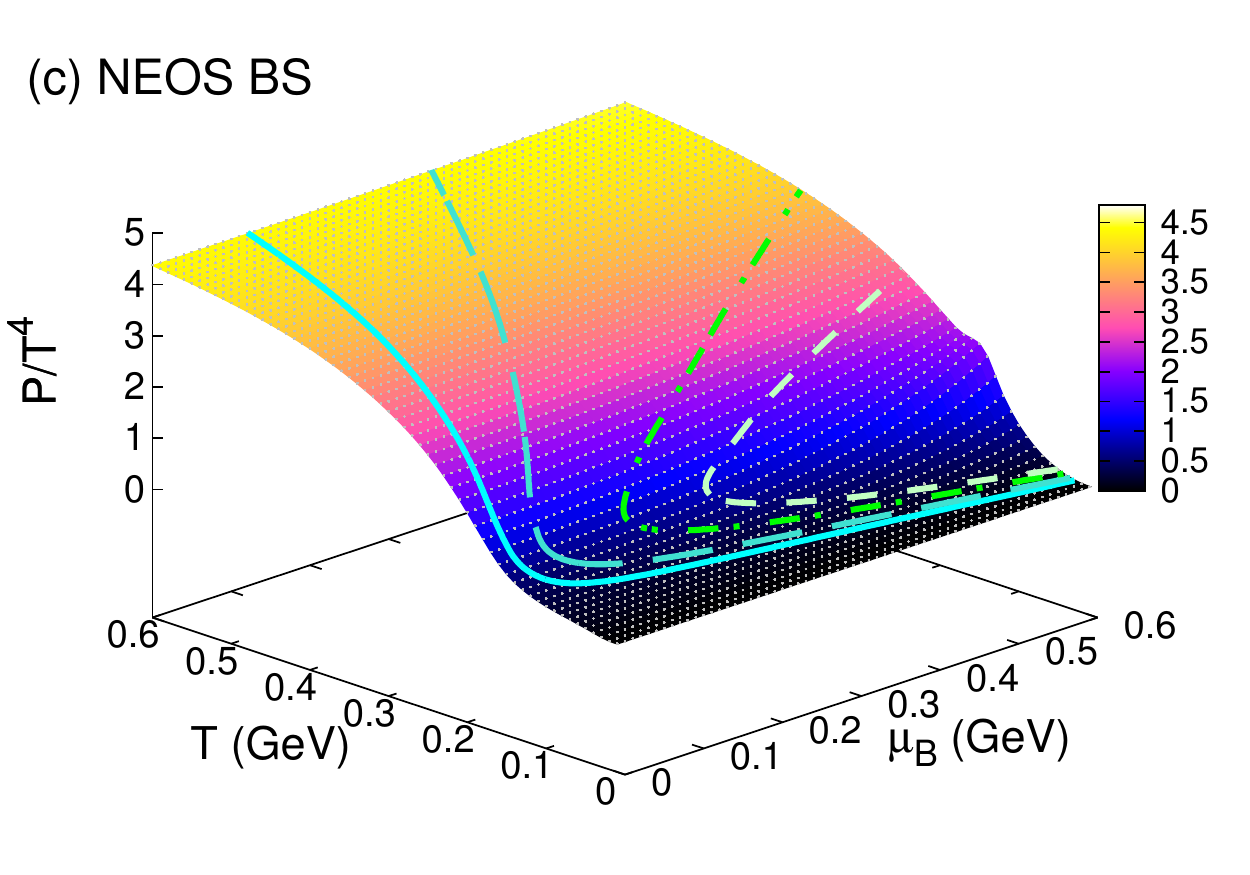}
\includegraphics[width=2.4in,bb=0 0 360 252]{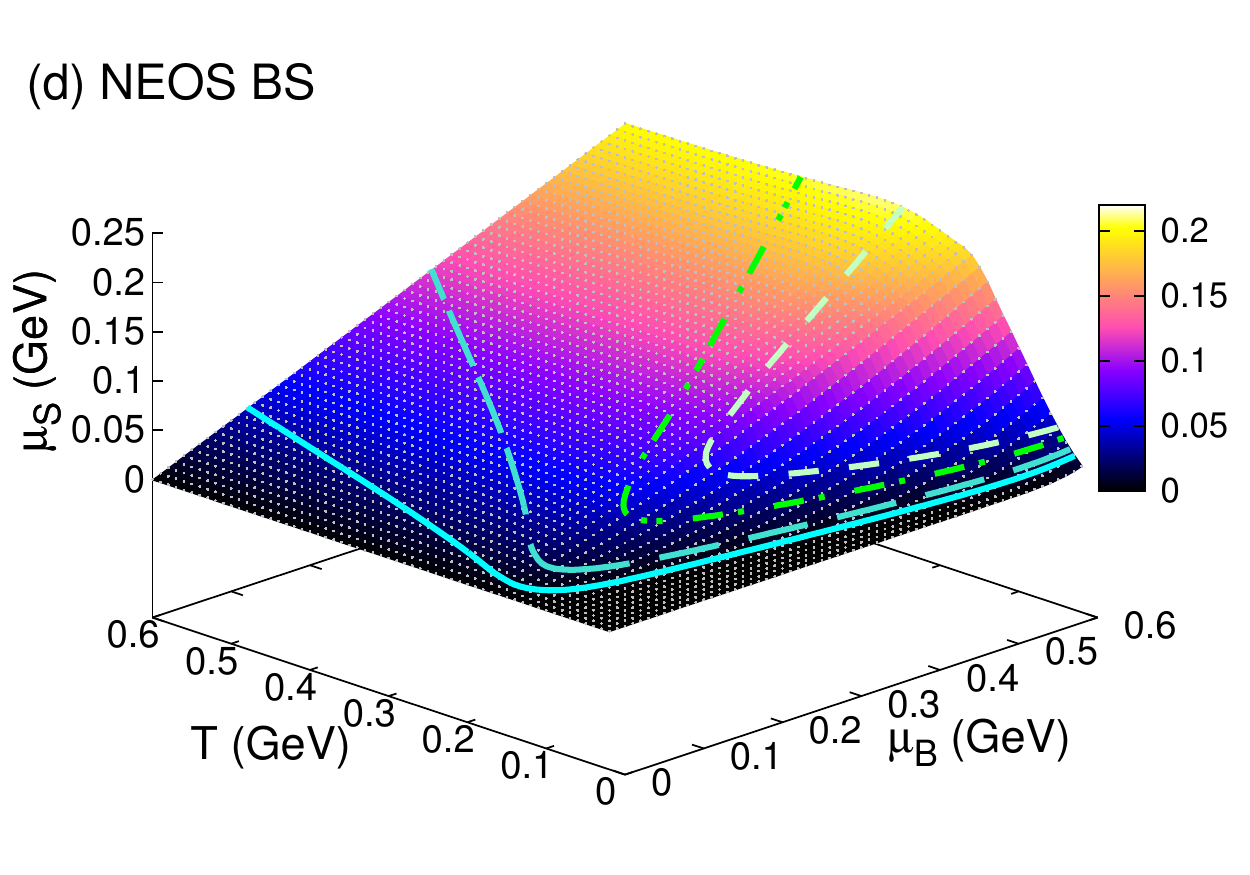}}
\centerline{\includegraphics[width=2.4in,bb=0 0 360 252]{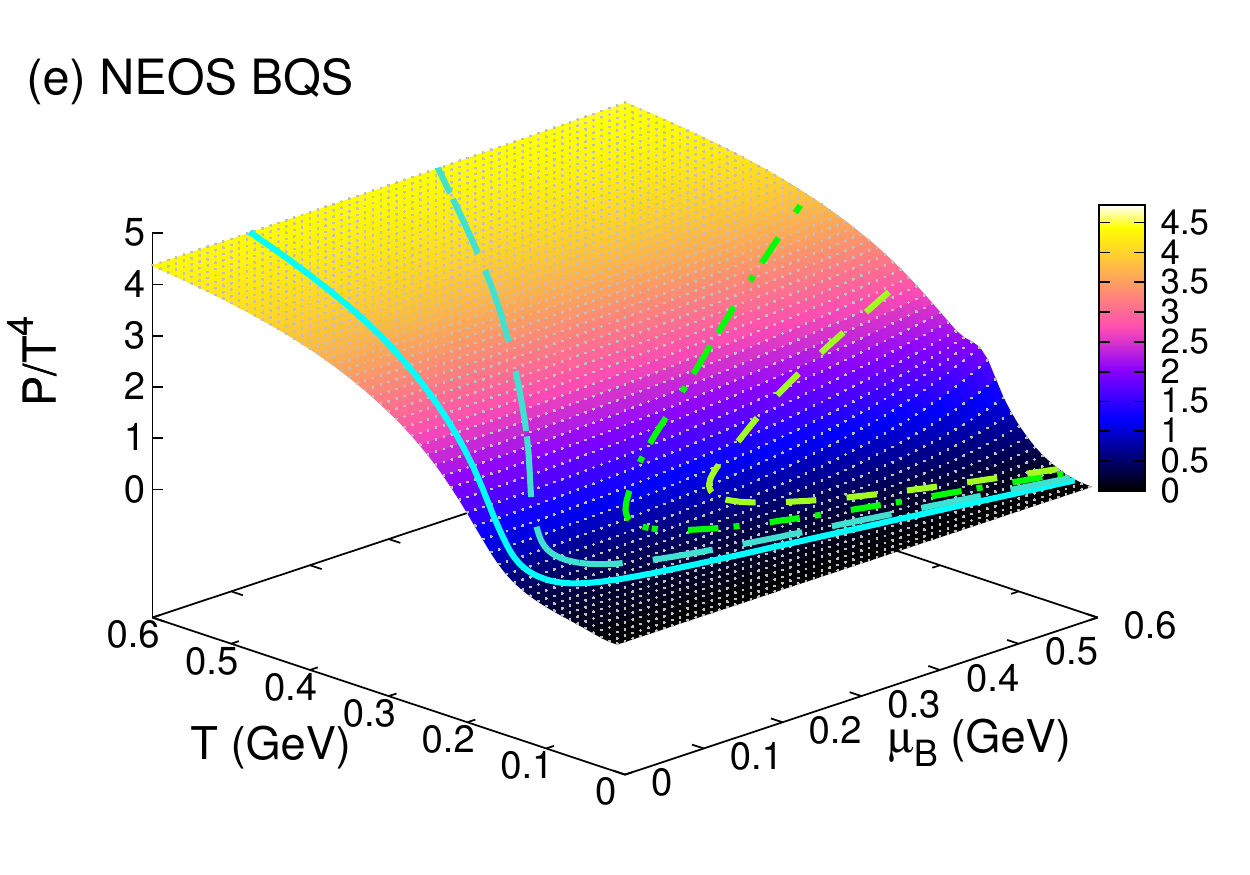}
\includegraphics[width=2.4in,bb=0 0 360 252]{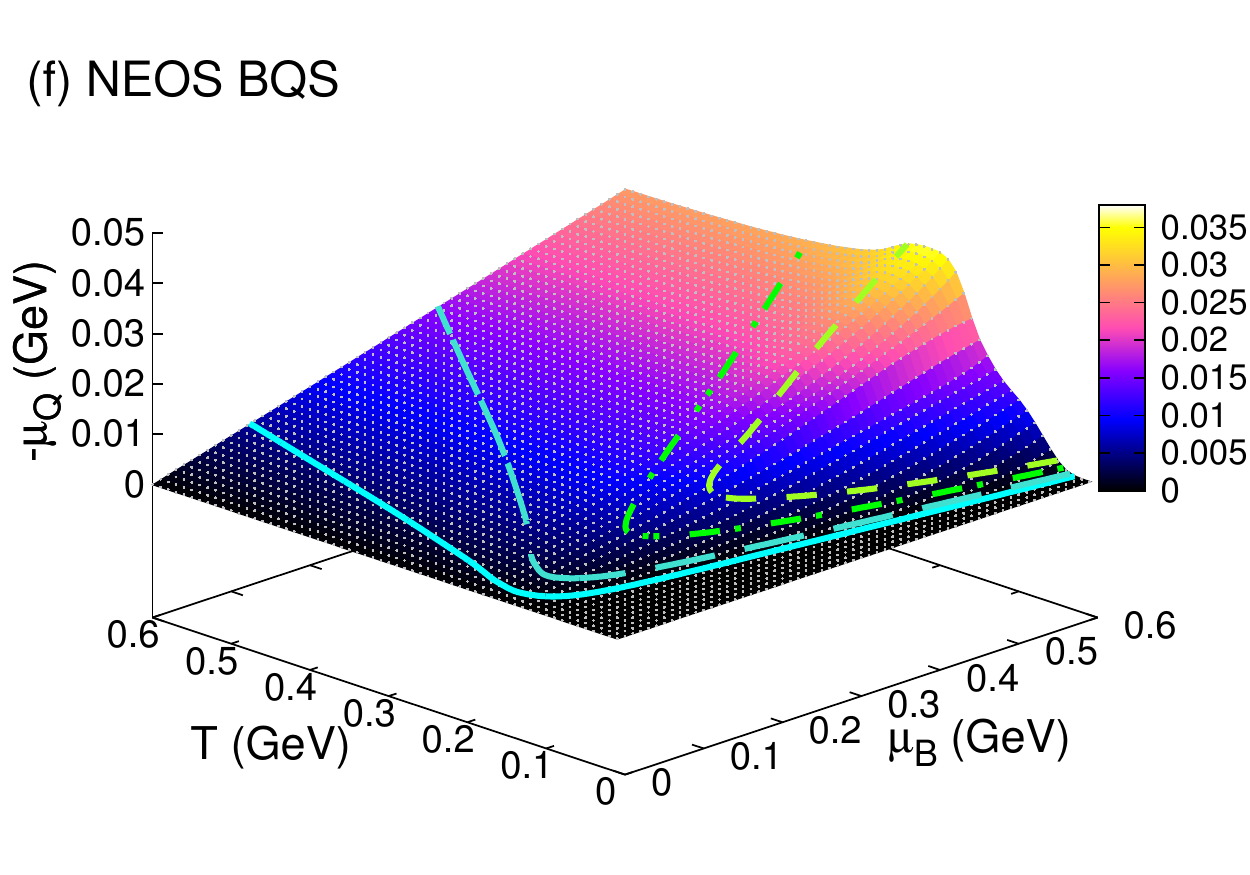}}
\caption{(a) The dimensionless pressure $P/T^4$ and (b) the dimensionless strangeness density $- n_S/T^3$ of \textsc{neos} B, (c) the dimensionless pressure $P/T^4$ and (d) the strangeness chemical potential $\mu_S$ of \textsc{neos} BS, and (e) the dimensionless pressure $P/T^4$ and (f) the electric charge chemical potential $-\mu_Q$ of \textsc{neos} BQS as functions of $T$ and $\mu_B$ \cite{Monnai:2019hkn}. The solid, long-dashed, dash-dotted, and short-dashed lines indicate the constant $s/n_B$ trajectories at 420, 144, 51, and 30, respectively. \label{fig:eos}}
\end{figure}

The dimensionless pressure $P/T^4$ as a function of $T$ and $\mu_B$ is shown in Fig.~\ref{fig:eos} (a) where $\mu_S = \mu_Q = 0$ (\textsc{neos} B). The trajectories of the constant entropy density to net baryon density ratio $s/n_B$ indicate the typical trajectory in the $T$-$\mu_B$ plane explored by collider experiments at each center-of-mass energy, because the net baryon density and -- in the ideal hydrodynamic approximation -- entropy density are conserved during the hydrodynamic evolution. $s/n_B = 420, 144, 51$, and $30$ correspond to $\sqrt{s_{NN}}=200, 62.4, 19.6$, and $14.5$ GeV \cite{Gunther:2016vcp}, respectively. 
It should be noted that there will be a range of $s/n_B$ for every collision since the medium is spatially inhomogeneous. Also, event-by-event fluctuations further smear the trajectories on the phase diagram.
$\mu_B/T$ is fixed on those trajectories when $s\sim T^3$ and $n_B \sim \mu_B T^2$ in the QGP phase. Once the trajectories enter the hadronic phase, they are bent toward larger $\mu_B$ because protons, the lightest baryons, are considerably heavier than pions. While this situation leads to a thermodynamically consistent crossover equation of state, it does not reflect the situation in nuclear collisions because the strangeness neutrality condition is violated as shown in Fig.~\ref{fig:eos} (b). The negative strangeness density is consistent with the expectation that positive $\mu_B$ leads to a system with more $s$ quarks and fewer $\bar{s}$ quarks. It approaches zero on the low temperature side because the lightest hadrons with strangeness are kaons, whose mass is non-negligible in the hadronic phase.

Once the strangeness neutrality condition $n_S=0$ is imposed, the pressure is meaningfully modified in the region where the $\mu_B/T$ is relatively large, as demonstrated in Fig.~\ref{fig:eos} (c). Figure~\ref{fig:eos} (d) shows that the strangeness neutrality condition leads to positive strangeness chemical potentials. The trajectories are shifted to the larger $\mu_B$ side by about 50\% in the QGP phase, because only $u$ and $d$ quarks contribute to $n_B$ in \textsc{neos} BS instead of $u, d$, and $s$ quarks in \textsc{neos} B, because in strangeness neutral systems strange quarks and antiquarks do not contribute to the net baryon number. The larger values of $\mu_B$ can be important in the hydrodynamic model because baryon diffusion, which is primarily driven by the spatial gradient of $\mu_B/T$ \cite{Monnai:2012jc,Li:2018fow,Denicol:2018wdp}, would be enhanced. The differences of the trajectories in \textsc{neos} B and BS are smaller in the hadronic phase because, as mentioned earlier, the lightest hadron with net baryon number and strangeness is the $\Lambda$ baryon, which is already heavy compared with the medium temperature. 

\begin{figure}[tb]
\centerline{\includegraphics[width=2.6in,bb=0 0 288 158]{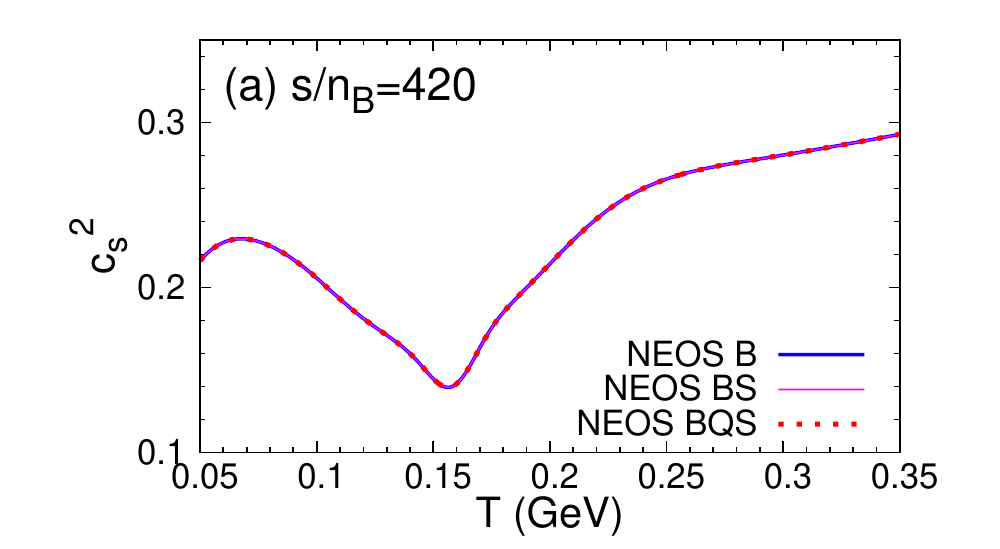}
\includegraphics[width=2.6in,bb=0 0 288 158]{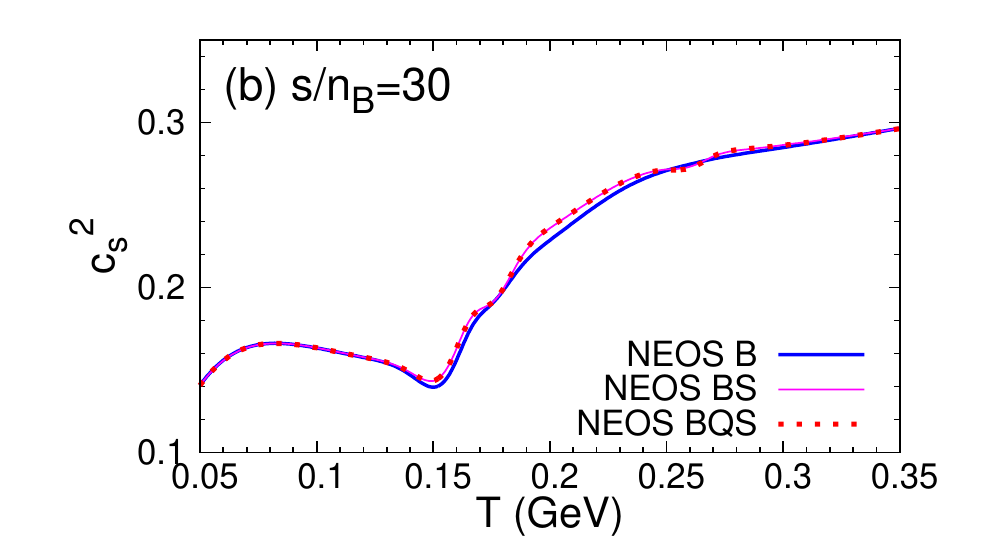}}
\caption{The thick solid, thin solid, and thick dotted lines are the sound velocity squared of \textsc{neos} B, BS, and BQS, respectively, as a function of temperature along the trajectories of (a) $s/n_B = 420$ and (b) $s/n_B = 30$. \cite{Monnai:2019hkn}.\label{fig:sv}}
\end{figure}

Finally, we study the situation of matter with fixed electric charge-to-baryon ratio $n_Q/n_B = 0.4$ and strangeness neutrality. Shown in Fig.~\ref{fig:eos} (e) and (f) are the dimensionless pressure and electric chemical potential, respectively. The pressure does not change much going from \textsc{neos} BS to BQS and neither do the trajectories, because $\mu_Q = 0$ implies $n_Q/n_B \sim 0.5$ which happens to be not too far from the more realistic situation. The negative electric chemical potential, nevertheless, is important in heavy-ion phenomenology, as it presents a quantitative explanation for the abundance of negative pions over positive pions observed in the experiments \cite{Alt:2007aa,Afanasiev:2002mx,Adamczyk:2017iwn}.

Figure~\ref{fig:sv} shows the sound velocity as a function of temperature for \textsc{neos} B, BS, and BQS. The sound velocity has a minimum because the pressure does not change significantly as a function of the energy density in the vicinity of the quark-hadron crossover (Eq.~\ref{eq:cs2}). Comparing the low baryon density ($s/n_B = 420$) and high baryon density ($s/n_B = 30$) results, the sound velocity in the hadronic phase is found to be suppressed and the minimum is shifted toward the lower temperature side for larger densities by a few MeV. The strangeness neutrality condition slightly increases the sound velocity, while the realistic electric charge-to-baryon ratio leads to negligible change. The quantity approaches the Stefan-Boltzmann limit $c_s^2 = 1/3$ at high temperatures, it reaches 94.8\% at $T=0.6$ GeV and 97.2\% at $T=0.8$ GeV of the limit for $s/n_B = 420$.

\begin{figure}[tb]
\centerline{\includegraphics[width=2.4in,bb=0 0 360 252]{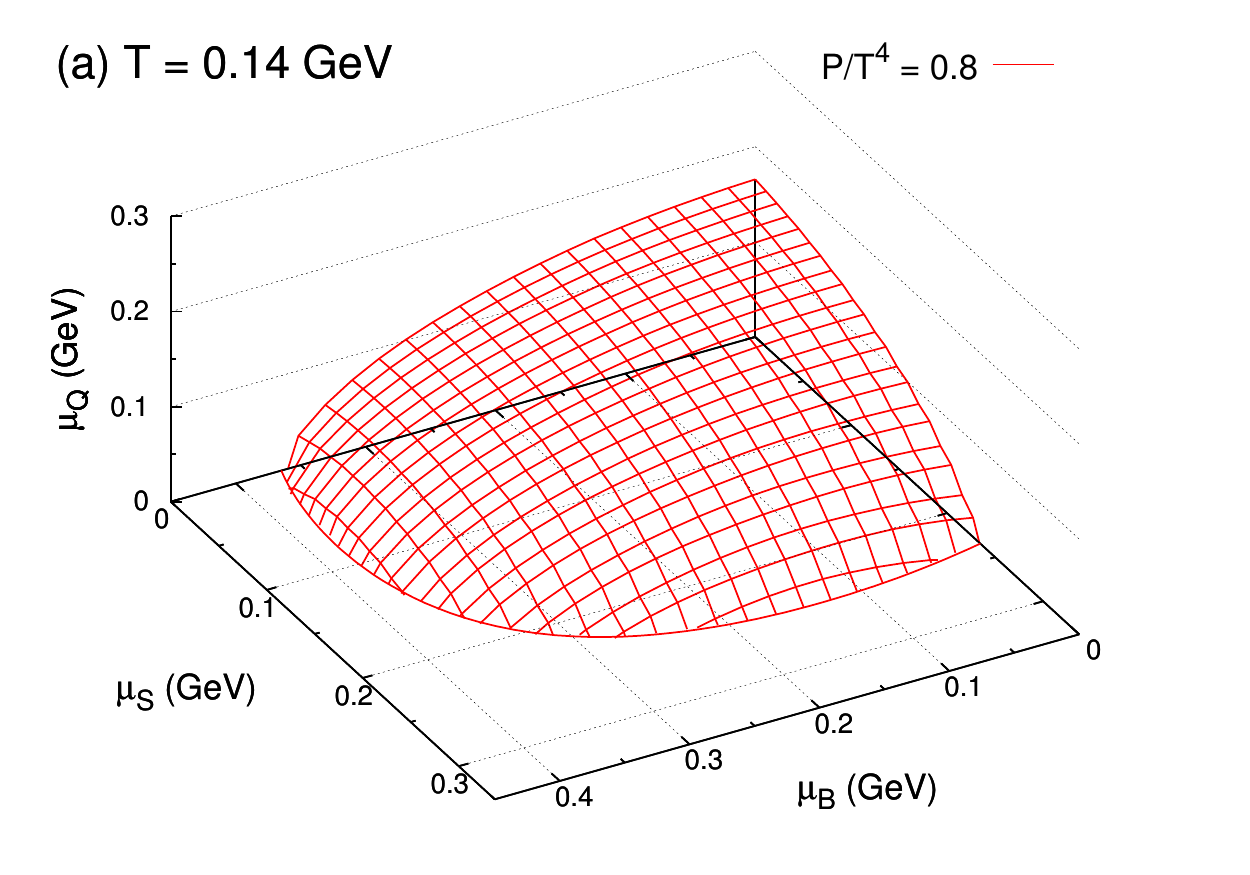}
\includegraphics[width=2.4in,bb=0 0 360 252]{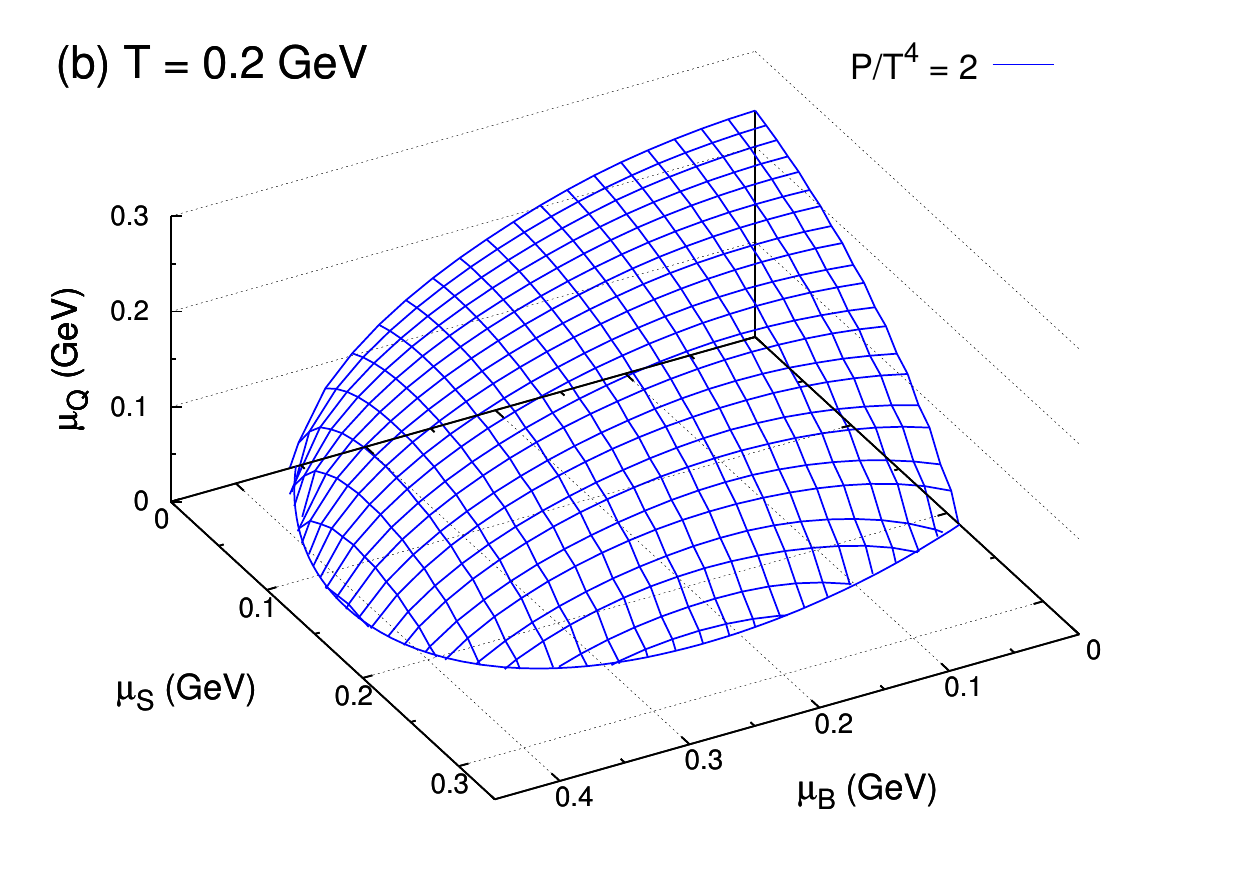}}
\caption{Isopressure planes in the chemical potential space in (a) the hadronic phase where $P/T^4 = 0.8$ and $T=0.14$ GeV and (b) the QGP phase where $P/T^4 = 2$ and $T=0.2$ GeV \cite{Monnai:2019hkn}.\label{fig:isopre}}
\end{figure}

The isopressure surface at constant temperatures in the chemical potential space is investigated to illustrate the interplay of multiple conserved charges. The numerical result in the hadronic phase where $P/T^4 = 0.8$ and $T=0.14$ GeV is shown in Fig.~\ref{fig:isopre}~(a). The intercepts can be defined as $P(\mu_B^\mathrm{int},0,0)=P(0,\mu_Q^\mathrm{int},0)=P(0,0,\mu_S^\mathrm{int})$. They are ordered as $\mu_B^\mathrm{int} > \mu_S^\mathrm{int}> \mu_Q^\mathrm{int}$, reflecting the mass ordering of the lightest hadrons to carry the respective charges, $m_p > m_K > m_\pi$. The situation is different in the QGP phase as shown in Fig.~\ref{fig:isopre}~(b) where $P/T^4 = 2$ and $T=0.2$ GeV. The intercept ordering $\mu_B^\mathrm{int} > \mu_Q^\mathrm{int}> \mu_S^\mathrm{int}$ is consistent with a parton gas interpretation that $\mu_B^\mathrm{int}/3\sim 2\mu_Q^\mathrm{int}/3 \sim \mu_S^\mathrm{int}$, though $\mu_S^\mathrm{int}$ is not as small owing to the fact that the strange quark mass is not negligible at the chosen temperature.

\begin{figure}[tb]
\centerline{\includegraphics[width=2.6in,bb=0 0 360 252]{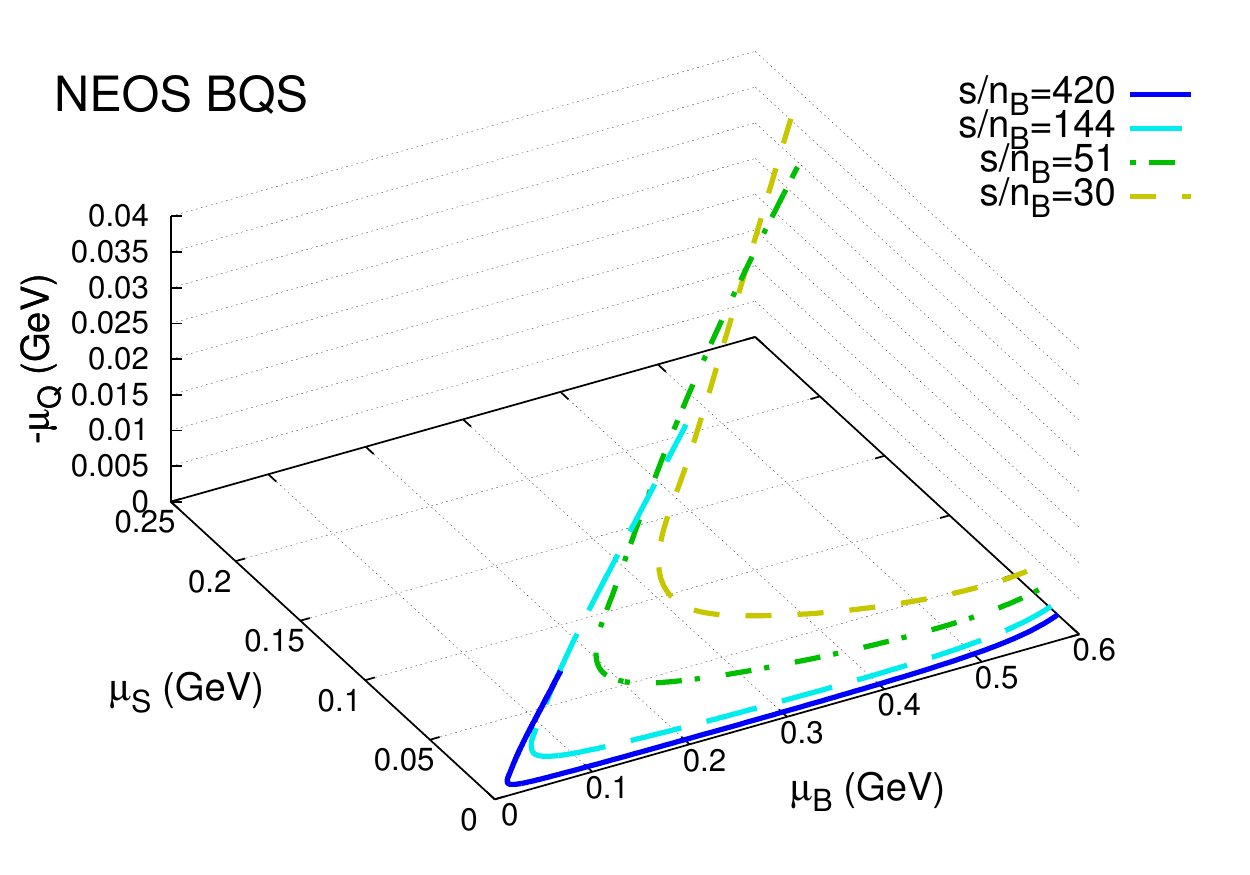}}
\caption{The solid, long-dashed, dash-dotted, and short-dashed lines indicate the constant $s/n_B$ trajectories at 420, 144, 51, and 30, respectively, in the $\mu_B$-$\mu_S$-$\mu_Q$ space \cite{Monnai:2019hkn}.\label{fig:multi}}
\end{figure}

The constant $s/n_B$ trajectories of \textsc{neos} BQS are plotted in the chemical potential space to illustrate typical regions explored in nuclear collisions (Fig.~\ref{fig:multi}). The end with larger values of $\mu_S$ and $|\mu_Q|$ corresponds to the high temperature region. One can see that the trajectories form a straight line in the QGP phase, because of the constraints $n_S = 0$ and $n_Q = 0.4 n_B$ under the leading order approximation of the partonic results \cite{Monnai:2019hkn}
\begin{eqnarray}
\begin{pmatrix} 
n_B \\ n_Q \\ n_S
\end{pmatrix}
= T^2
\begin{pmatrix} 
\chi_2^B & \chi_{1,1}^{B,Q} & \chi_{1,1}^{B,S} \\ \chi_{1,1}^{B,Q} & \chi_2^Q & \chi_{1,1}^{Q,S} \\ \chi_{1,1}^{B,S} & \chi_{1,1}^{Q,S} & \chi_2^S
\end{pmatrix}
\begin{pmatrix} 
\mu_B \\ \mu_Q \\ \mu_S
\end{pmatrix},
\end{eqnarray}
which leads to $\mu_B = 4.6 n_B/T^2$, $\mu_Q = - 0.2 n_B/T^2$, and $\mu_S = 1.6 n_B/T^2$. As mentioned earlier, they deviate from the straight line in the hadronic phase in the direction of larger baryon chemical potentials because of the mass difference between protons and pions. The second bend towards larger strangeness chemical potential near the low temperature end is induced by the mass difference between kaons and pions. It is important to note that one does not explore the $T$-$\mu_B$ plane but the $T$-$\mu_B$-$\mu_Q$-$\mu_S$ space in nuclear collider experiments. This should be taken into account when analyzing the experimental data to learn about the phase structure of QCD.

\subsection{Applications to nuclear collisions}
The phenomenological consequences of the conditions of strangeness neutrality and electric charge-to-baryon ratio of heavy nuclei are studied by using the hydrodynamic model \cite{Schenke:2019ruo} of relativistic nuclear collisions. We consider Pb+Pb collisions at $\sqrt{s_{NN}}=17.3$~GeV as conducted at the CERN SPS \cite{Afanasiev:2002mx,Alt:2004kq,Alt:2006dk,Alt:2007aa,Alt:2008qm,Alt:2008iv}.

The initial conditions for the hydrodynamic model are calculated using an event-by-event dynamical Glauber model \cite{Shen:2017bsr}. The Glauber model is a framework that provides initial geometrical configurations in the transverse plane based on the Woods-Saxon potential and inelastic nucleon-nucleon cross section \cite{Miller:2007ri}. Its improved version, the Glauber-Lexus model \cite{Monnai:2015sca}, takes into account the exchange of longitudinal momentum \cite{Jeon:1997bp}. The dynamical Glauber model is the four-dimensional version in the sense that the energy and net baryon number densities are introduced to the system as each sub-collision of target and projectile nucleons occurs over time.

The numerical implementation \textsc{music} \cite{Schenke:2010nt,Schenke:2010rr,Schenke:2011bn} is used to perform the three-dimensional hydrodynamic simulation. A simple choice of transport coefficients, namely a shear viscosity of $\eta/s = 0.08$ and vanishing bulk viscosity and baryon diffusion, is employed to minimize ambiguities. Particlization is assumed to occur on a surface of constant energy density, defined by the switching energy density $e_\textrm{sw}$. Particles are then further evolved using the hadron cascade model Ultra-relativistic Quantum Molecular Dynamics (UrQMD) \cite{Bass:1998ca,Bleicher:1999xi}.

\begin{figure}[tb]
\centerline{\includegraphics[width=2.4in,bb=0 0 576 432]{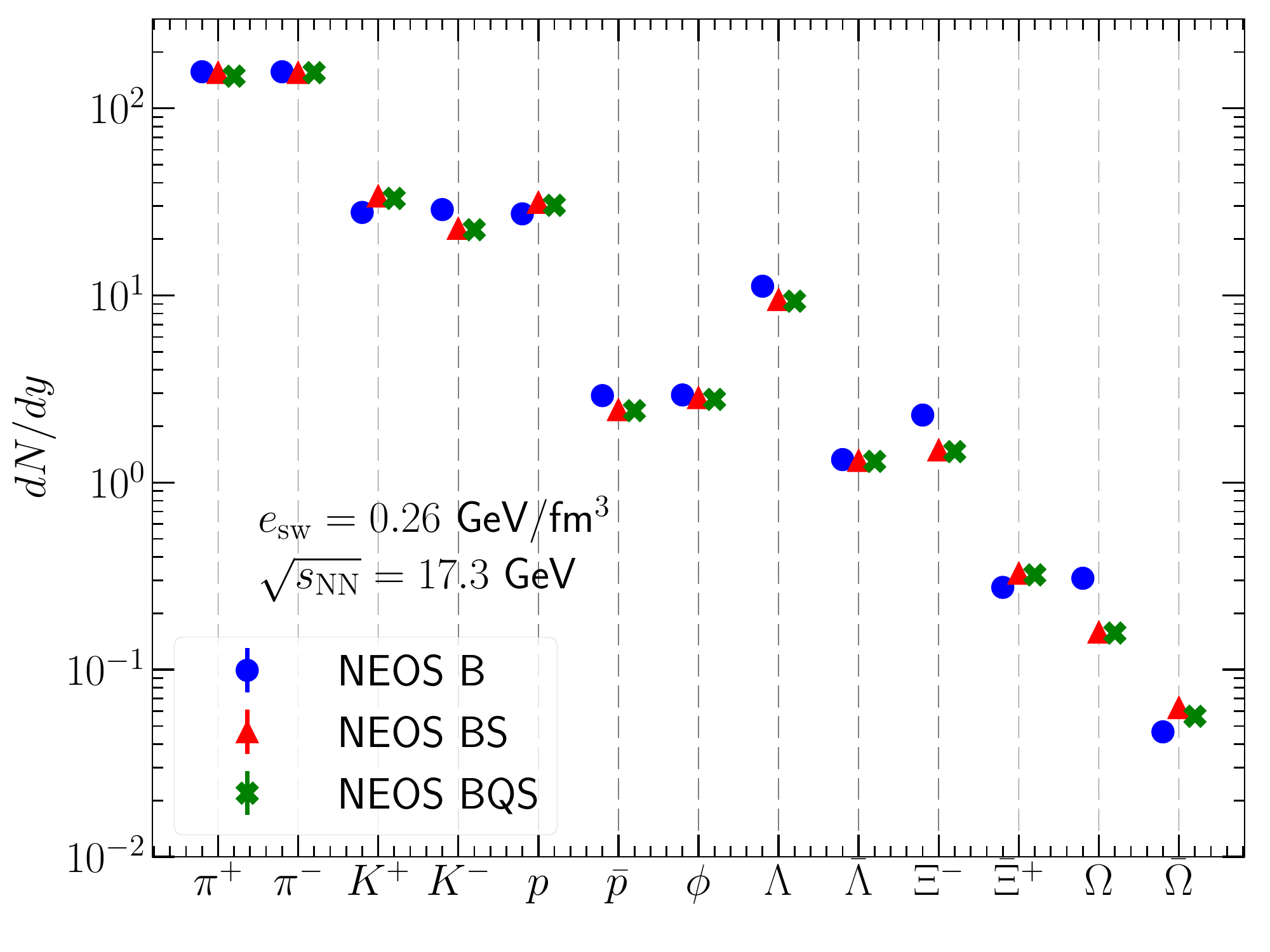}
\includegraphics[width=2.4in,bb=0 0 576 432]{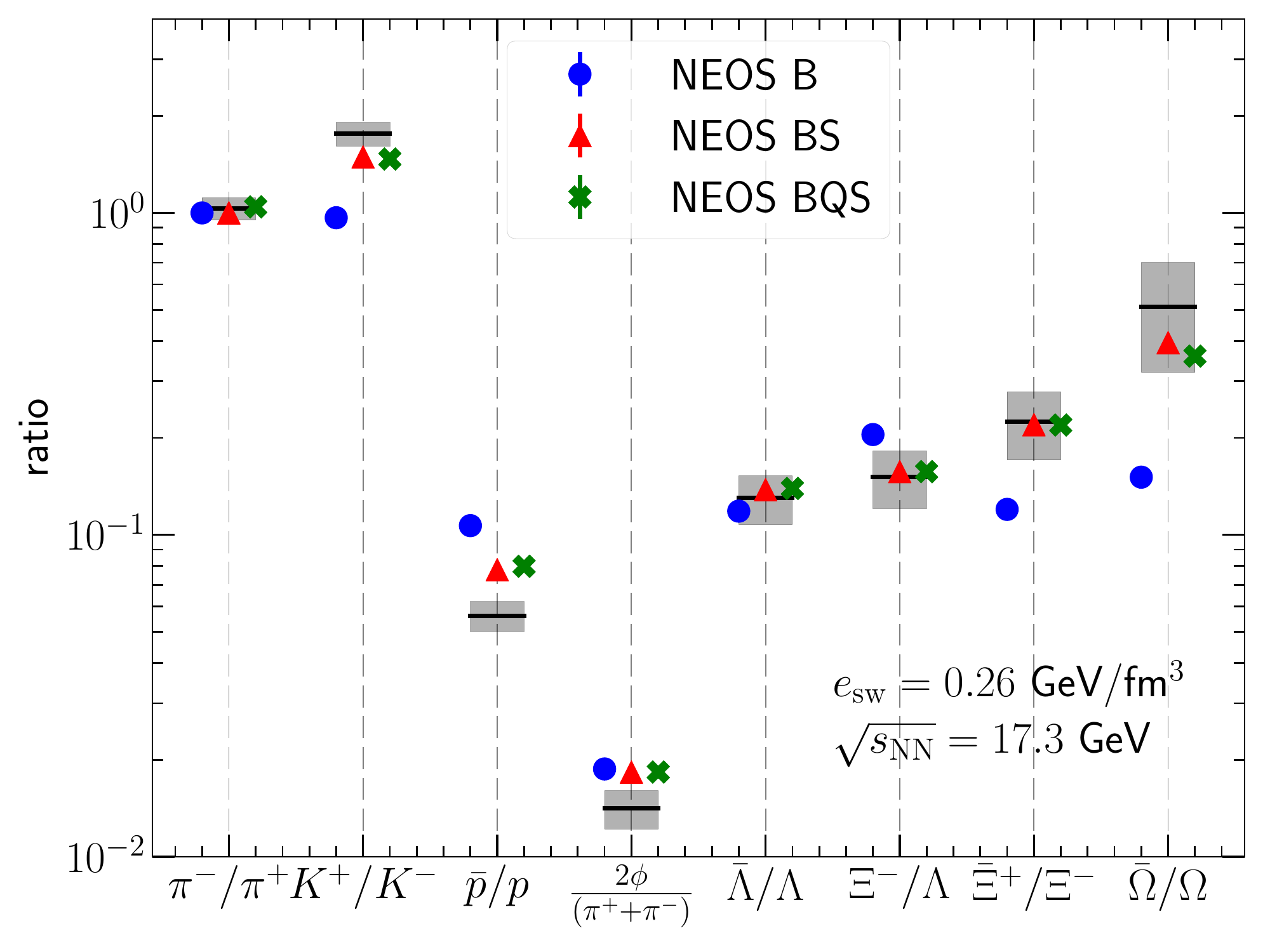}}
\caption{(Left) The hadronic yields for the most central Pb+Pb collisions at $\sqrt{s_{NN}}=17.3$~GeV calculated with \textsc{neos} B, BS, and BQS represented by circular, triangular and cross symbols, respectively. \cite{Monnai:2019hkn}. (Right) The particle-antiparticle ratios with the same conditions compared with the experimental data \cite{NA49data}. \label{fig:yields}}
\end{figure}

\begin{table}[tb]
\tbl{List of hadronic chemical potentials.}
{\begin{tabular}{cc}
\toprule
Hadrons & Chemical potentials \\ \midrule
$\pi^+$ & $\mu_Q$ \\
$K^+$ & $\mu_Q+\mu_S$ \\
$\phi$ & 0 \\
$p$ & $\mu_B+\mu_Q$ \\
$\Lambda$ & $\mu_B-\mu_S$ \\
$\Xi^-$ & $\mu_B-\mu_Q-2\mu_S$ \\
$\Omega$ & $\mu_B-\mu_Q-3\mu_S$ \\ \botrule
\end{tabular}
\label{table:muhad}}
\end{table}

The simulated yields of particles and antiparticles in most central events (using $e_\textrm{sw} = 0.26$ GeV/fm$^3$) are shown in Fig.~\ref{fig:yields}~(left) and their ratios in Fig.~\ref{fig:yields}~(right) for the three different versions of the equation of state. Comparison of \textsc{neos} B and BS results to the experimental data from SPS shows that the strangeness neutrality condition improves the description of the particle-antiparticle ratios of the hadrons with finite strangeness chemical potential, $K$, $\Lambda$, $\Xi$, and $\Omega$. The antiproton-proton ratio is also modified and moves closer to the data because of the aforementioned enhancement in the baryon chemical potential, when imposing strangeness neutrality. The differences between the \textsc{neos} BS and BQS results are rather small because the electric chemical potential is small for the collisions of heavy nuclei. As mentioned before, it is still phenomenologically important because it explains the experimental result that the anti-pion to pion ratio is greater than one. A list of particle species along with the chemical potentials that affect their respective yields is given in  Table~\ref{table:muhad}. 

\begin{figure}[tb]
\centerline{\includegraphics[width=2.4in,bb=0 0 576 432]{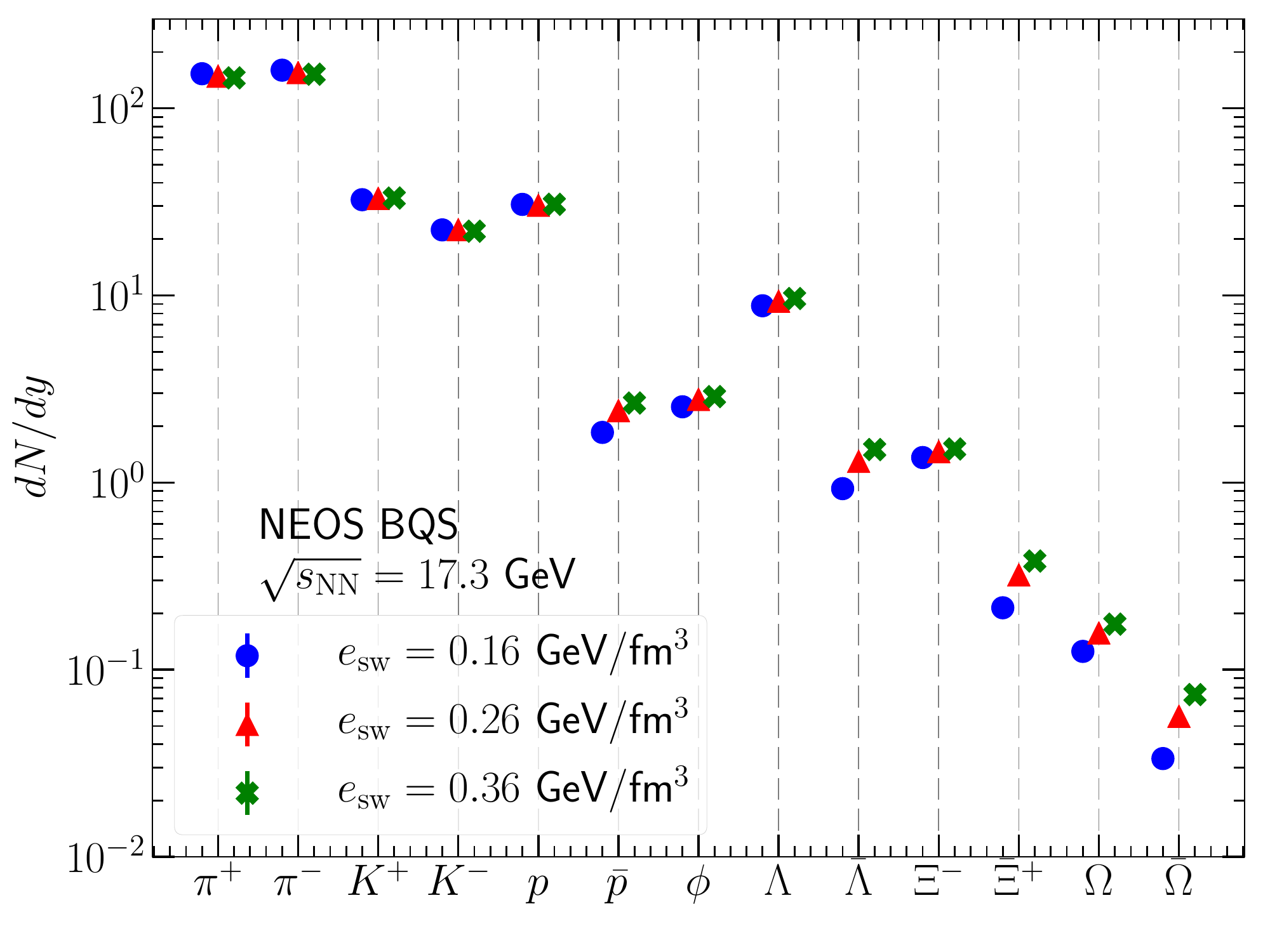}
\includegraphics[width=2.4in,bb=0 0 576 432]{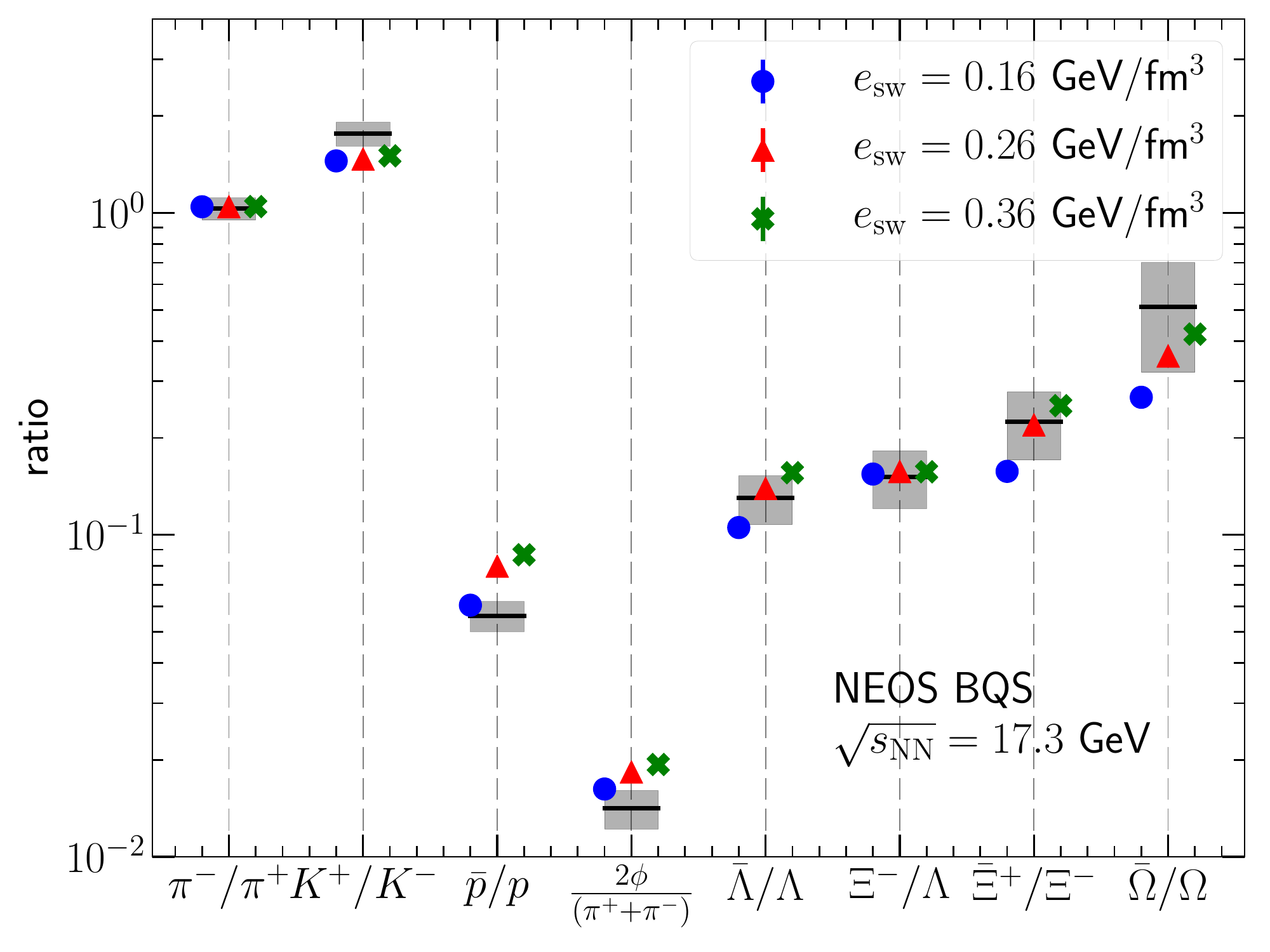}}
\caption{(Left) The hadronic yields for the most central Pb+Pb collisions at $\sqrt{s_{NN}}=17.3$~GeV estimated with \textsc{neos} BQS at $e_\textrm{sw} = 0.16, 0.26$, and $0.36$ GeV/fm$^3$ represented by circular, triangular and cross symbols, respectively. \cite{Monnai:2019hkn}. (Right) The particle-antiparticle ratios with the same conditions compared with experimental data \cite{NA49data}. \label{fig:ews}}
\end{figure}

In Fig.~\ref{fig:ews}, the switching energy density dependence is studied for particle yields (left) and ratios (right) using \textsc{neos} BQS. Chemical equilibrium is assumed down to lower temperatures when a lower switching energy density is considered. The results mostly agree with the experimental data when $e_\textrm{sw} = 0.16$-$0.36$ GeV/fm$^3$. The yields of antibaryons are most affected as $e_\textrm{sw}$ decreases, possibly because of the interplay of the enhancement of the baryon chemical potential at particlization (see Fig.\,\ref{fig:eos} (e)) and the suppression of heavier particle production in the thermal bath. The two effects tend to cancel for baryons while they are additive for antibaryons.

\section{Comparison of equation of state models}

We numerically compare different models of the QCD equation of state used in hydrodynamic simulations of relativistic nuclear collisions. Then effects of the differences on hydrodynamic evolution are investigated.

\subsection{Thermodynamic properties}

\begin{figure}[tb]
\centerline{
\includegraphics[width=2.4in,bb=0 0 576 432]{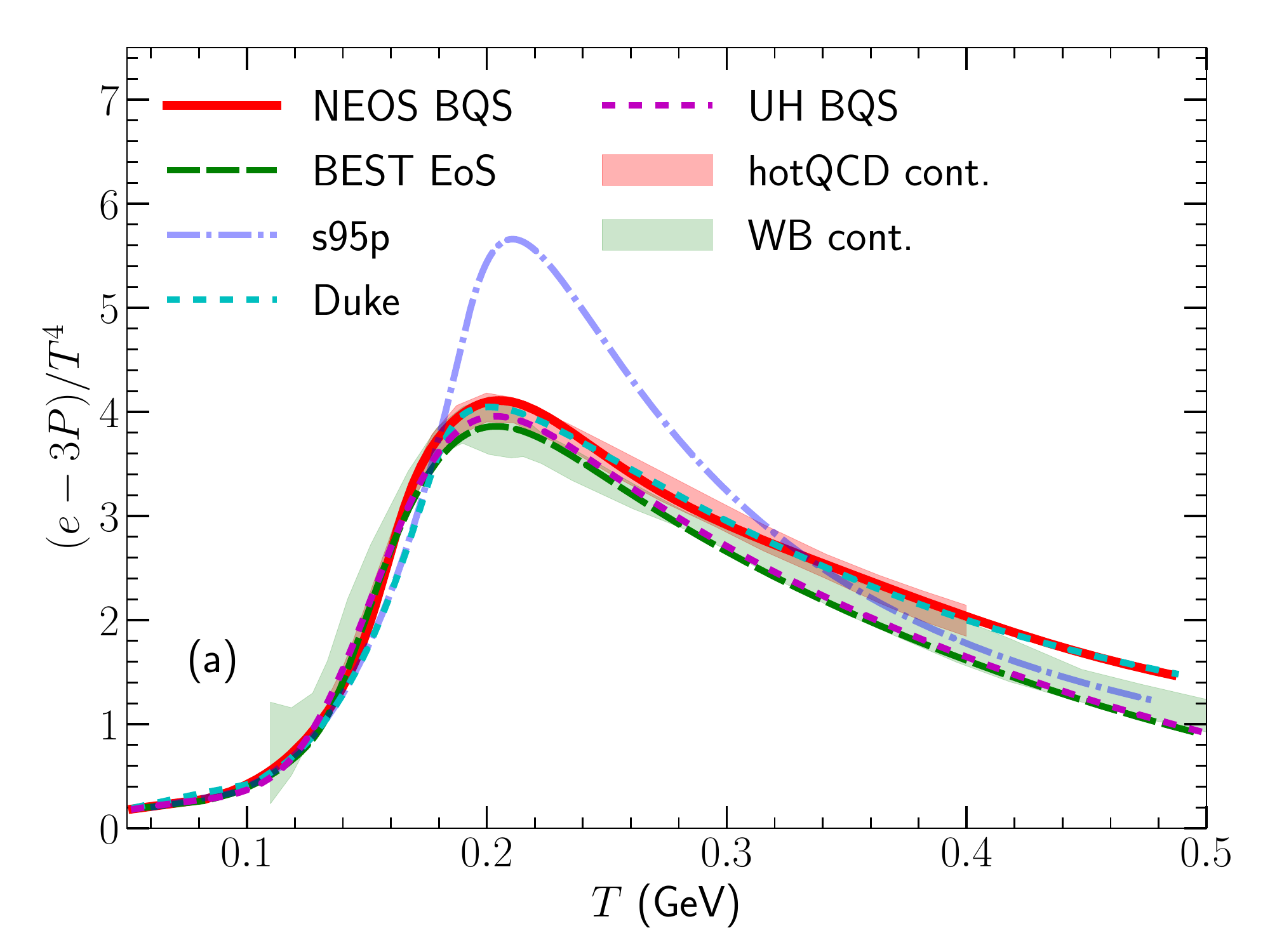}
\includegraphics[width=2.4in,bb=0 0 576 432]{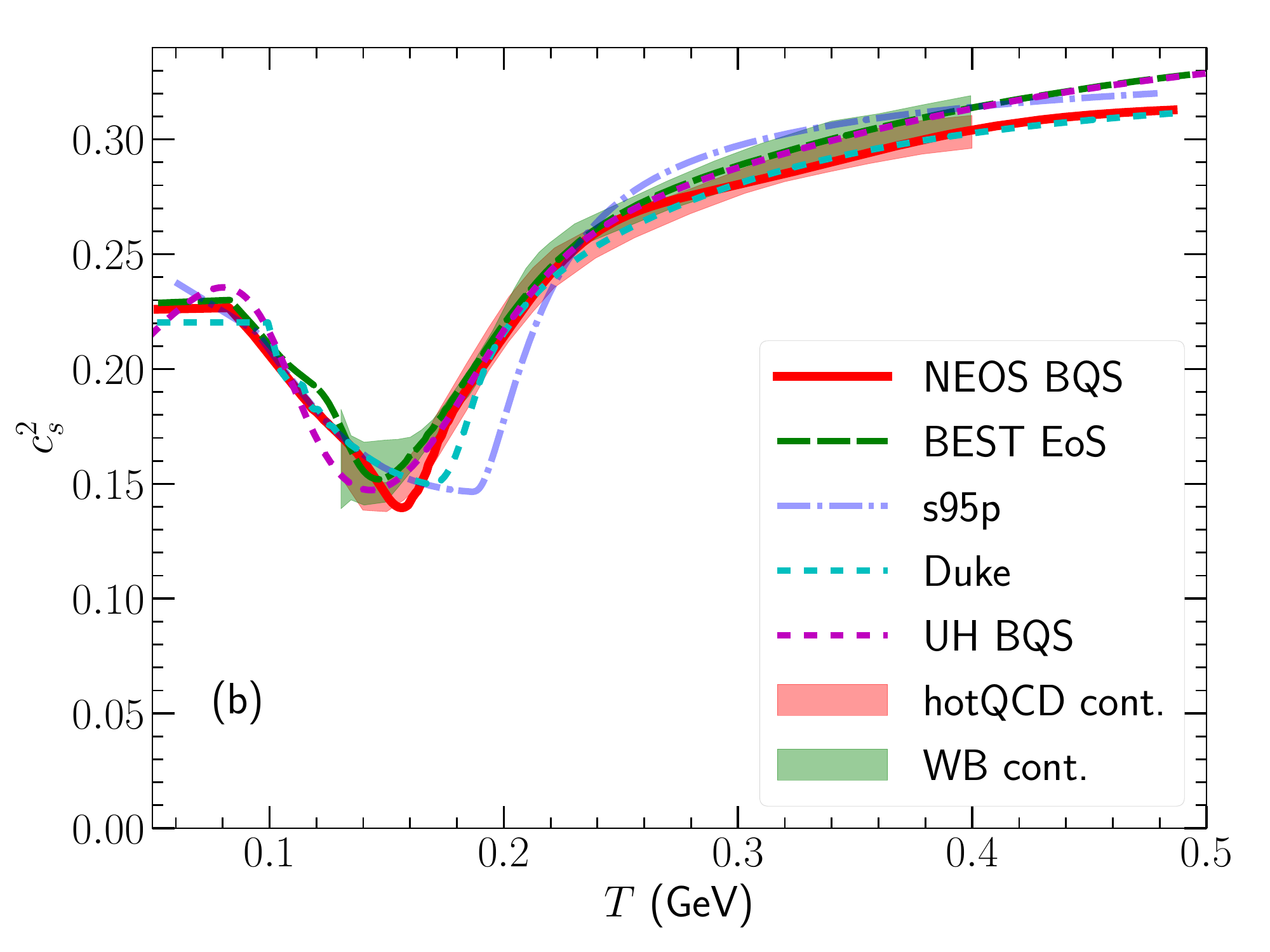}}
\caption{(a) Comparison of the trace anomalies from lattice QCD simulations and lattice QCD based equation of state models. (b) Comparison of the sound velocities extracted from the equation of state models and lattice QCD calculations.\label{fig:comp_zero}}
\end{figure}

Shown in Fig.~\ref{fig:comp_zero} (a) are the trace anomalies $(e-3P)/T^4$ from lattice QCD simulations and several equation of state models alongside \textsc{neos}, that we discussed in the previous section. \textsc{neos} and Duke \cite{Moreland:2015dvc} 
equations of state show agreement with the continuum limit result of the HotQCD Collaboration, on which their structure is based. Similarly, BEST \cite{Parotto:2018pwx} 
and University of Houston (denoted as UH) \cite{Noronha-Hostler:2019ayj} equations of state agree with the results from the Wuppertal-Budapest (WB) Collaboration, utilized for their construction. s95p-v1 (s95p) \cite{Huovinen:2009yb} is one of the earliest works on the hybrid equation of state and the deviation from the rest of the models may be owing in part to the difference in the lattice QCD data used. Its parametrization and matching procedure are also different owing to the now-resolved discrepancy between the resonance gas and early lattice data with non-physical pion mass. An updated version, employing more recent lattice QCD results, s83s$_{18}$, has recently been released \cite{Auvinen:2020mpc}. It is noteworthy that Duke and s95p results are very similar in the hadronic phase. Finally, we point out that the two shown lattice QCD results for the trace anomaly in the continuum limit are consistent.

The sound velocities are shown in Fig.~\ref{fig:comp_zero} (b). The basic structure of having a minimum of $c_s^2$ near the crossover is found in all models and lattice simulations. The exact location of the minimum is sensitive to the details of the construction of each model, such as the connecting temperature and width. Again, by construction \textsc{neos} and Duke equations of state agree with the HotQCD result -- and BEST and UH equations of state with the WB result -- at higher temperatures.

\begin{figure}[tb]
\centerline{
\includegraphics[width=2.4in,bb=0 0 576 432]{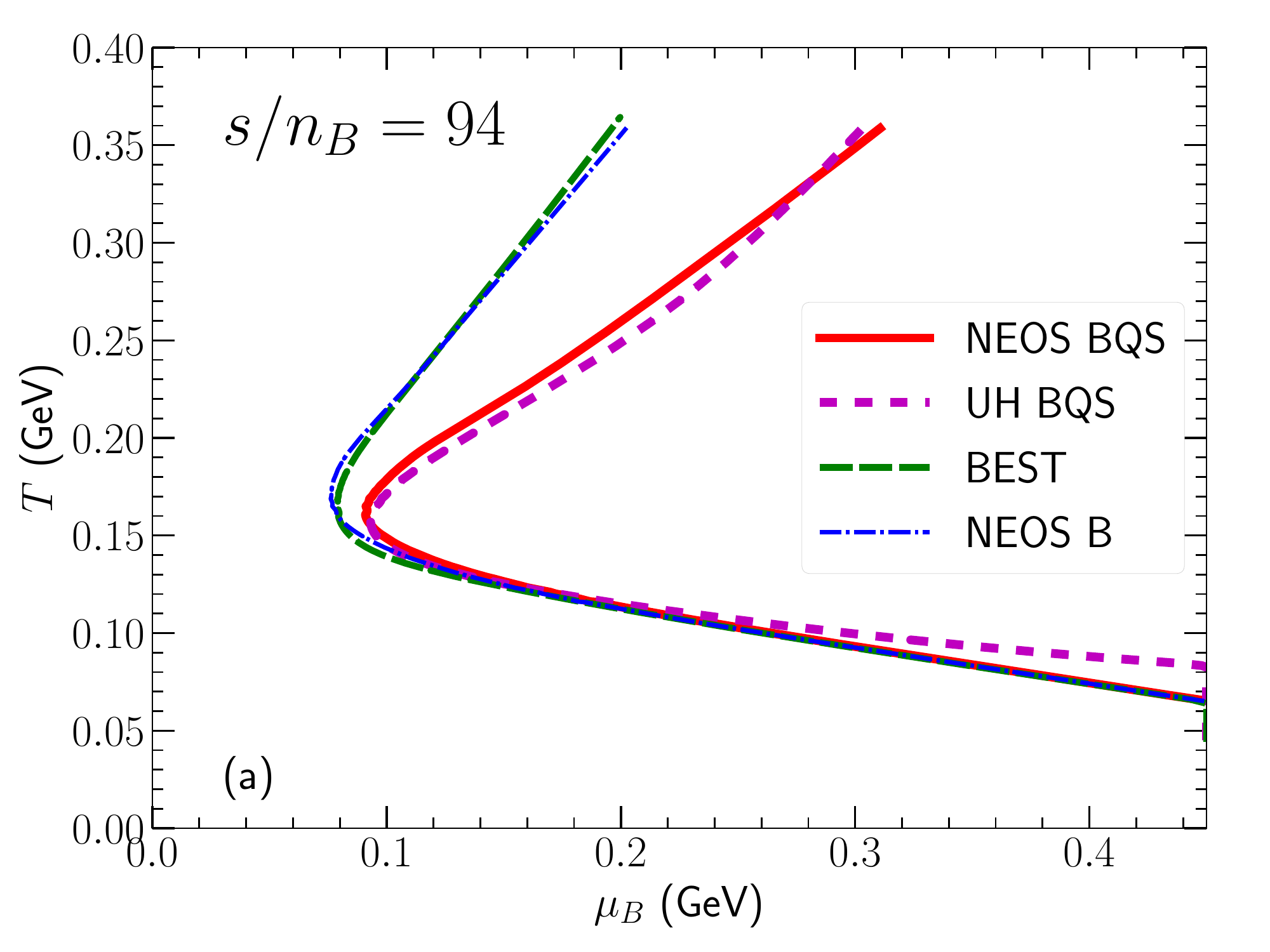}
\includegraphics[width=2.4in,bb=0 0 576 432]{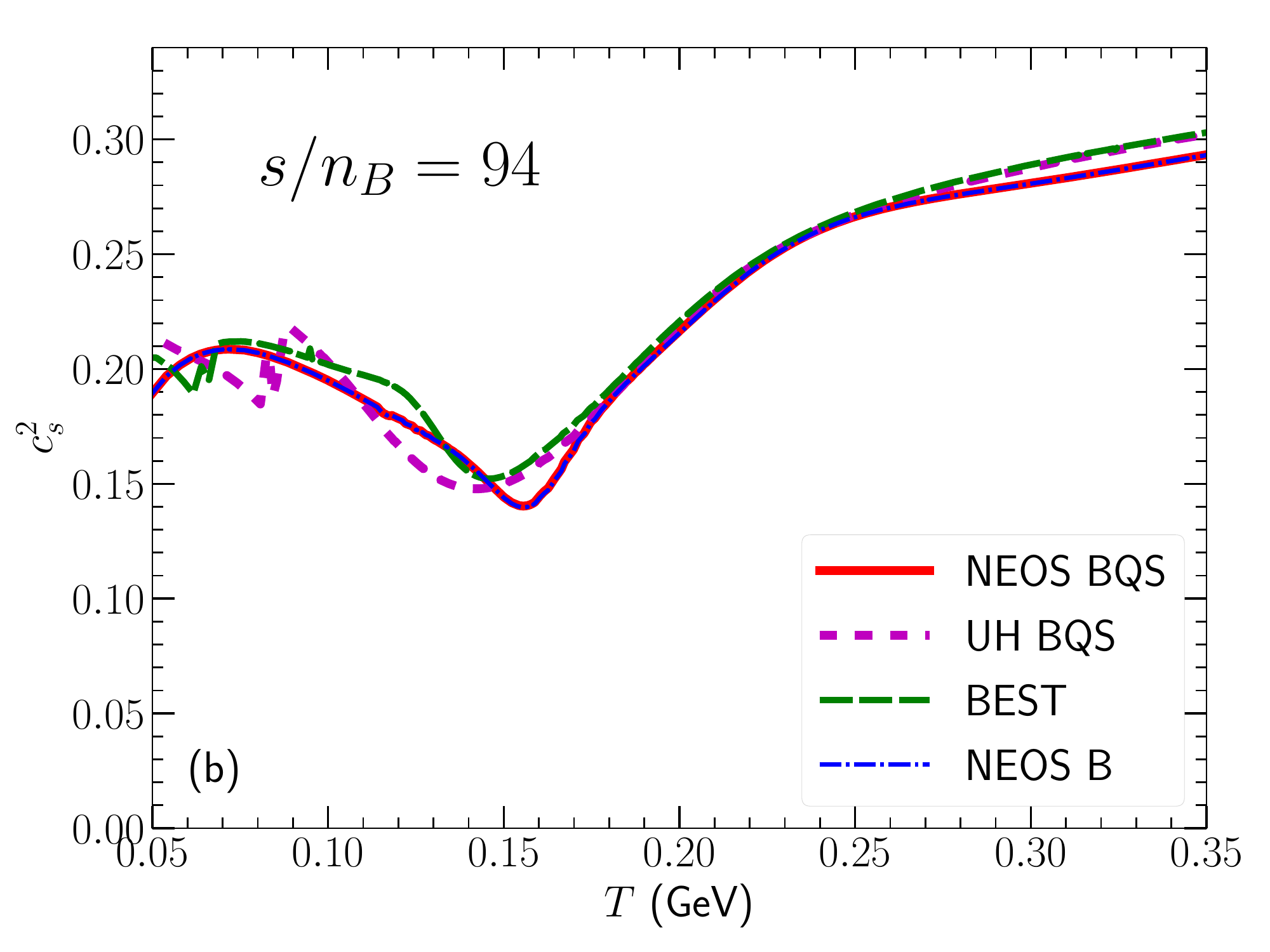}}
\caption{(a) Comparison of the phase trajectories for constant $s/n_B = 94$ from BEST equation of state vs. \textsc{neos} B and UH BQS vs. \textsc{neos} BQS. (b) Comparison of the sound velocities extracted from the equation of state models.\label{fig:comp_finitenB}}
\end{figure}

The comparison of the trajectories on the phase diagram for constant $s/n_B=94$, which approximately corresponds to the collision energy of $\sqrt{s_{NN}}=39$ GeV \cite{Ratti:2016lrh}, is shown in Fig.~\ref{fig:comp_finitenB} (a) to illustrate the properties of the equations of state at finite density. The phase trajectories for \textsc{neos} B and BEST as well as those for \textsc{neos} BQS and UH BQS behave similarly with small differences in the QGP phase, which may come from the difference in the lattice QCD data employed. The effect of additional charges to the baryon chemical potential is as discussed in the previous section. The difference between \textsc{neos} BQS and UH BQS results in the hadronic phase may come from the difference in the hadronic components used in the resonance gas \cite{Alba:2017mqu} and the structural difference that the UH BQS equation of state is expanded up to the second and fourth order in $\mu_{B,Q,S}/T$ in the hadronic phase to perform matching to the lattice data in the susceptibilities, whereas \textsc{neos} uses the hadron resonance gas without truncation as the matching is done for the pressure.

Figure~\ref{fig:comp_finitenB} (b) shows the sound velocities at finite densities on the $s/n_B = 94$ trajectories. One can see that they are sensitive to the details of the equation of state used. The results obtained with \textsc{neos} B and BQS are similar as previously observed in Fig.~\ref{fig:sv}. It should be noted that the small wiggles in UH BQS and BEST equations of state at low temperature are artifacts caused by a cut-off at $\mu_B = 0.45$\,GeV.

\subsection{Hydrodynamic evolution}

\begin{figure}[tb]
\centerline{
\includegraphics[width=2.4in,bb=0 0 576 432]{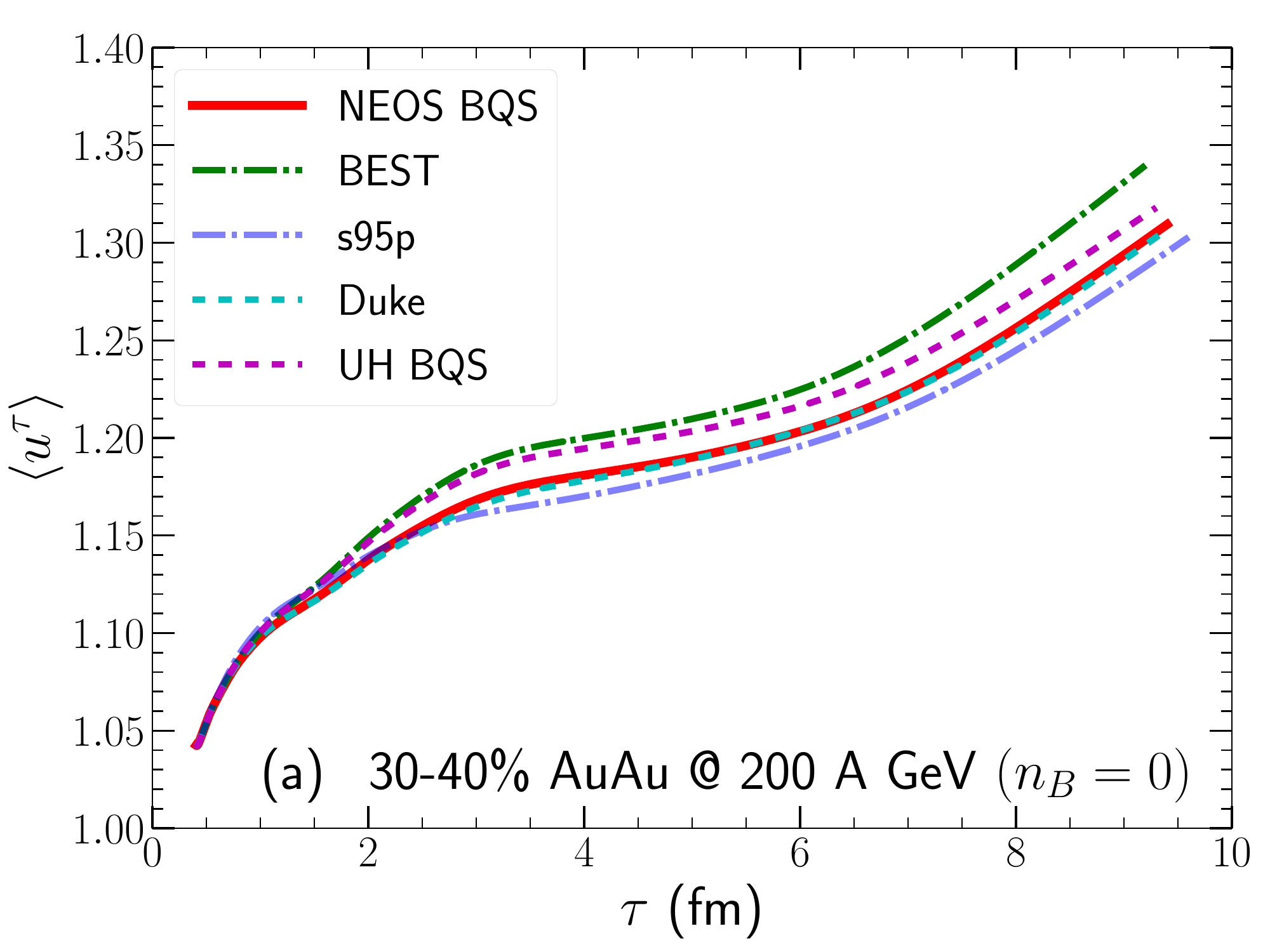}
\includegraphics[width=2.4in,bb=0 0 576 432]{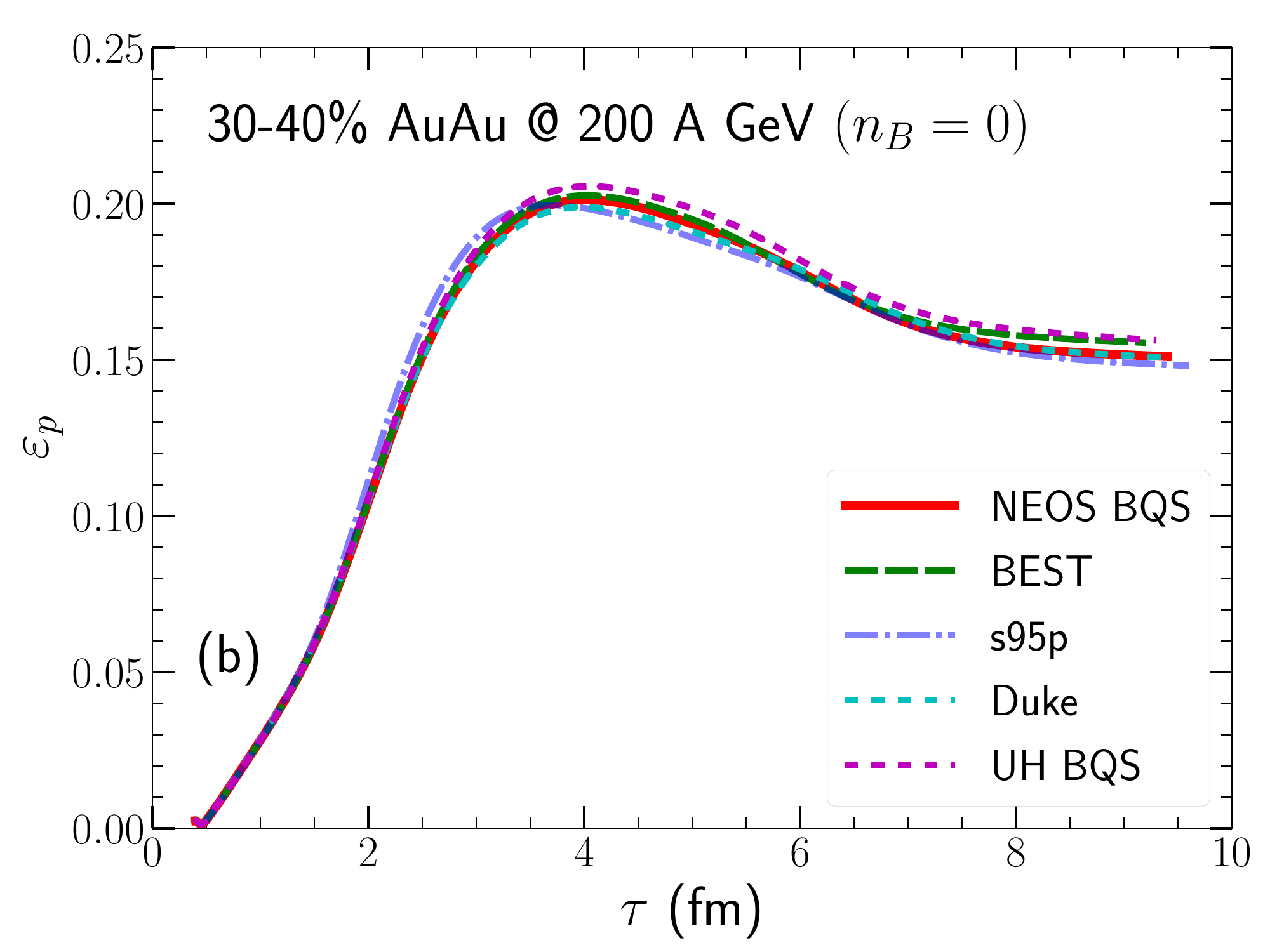}}
\caption{(a) The time evolution of averaged $u^\tau$ for an Au+Au collision in 30-40\% centrality at 200 GeV with different equations of state. (b) Similar comparison for the time evolution of the momentum anisotropy $\varepsilon_p$. \label{fig:comp_hydroevo_zeronB}}
\end{figure}

We now compare the hydrodynamic evolution in heavy ion collisions with different equations of state. Figure~\ref{fig:comp_hydroevo_zeronB} (a) shows the time-evolution of the average time-like flow component $\langle u^\tau \rangle$ in one 30-40\% Au+Au collision at $\sqrt{s_{NN}}=200$ GeV with the IP-Glasma initial condition \cite{Schenke:2012wb,Schenke:2020mbo}. The quantity is closely related to radial flow, which affects the slope of the particles' transverse momentum spectra ($u^\tau$ is closely related to the transverse flow velocity $u_\perp$ via the flow normalization condition $u \cdot u = 1$, particularly when neglecting longitudinal flow $u^\eta$). $\langle u^\tau \rangle$ increases 
with time and exhibits similar behavior in all cases. \textsc{neos} BQS and Duke equations of state lead to similar results.
The BEST and UH BQS equations of state lead to larger $\langle u^\tau \rangle$, while $\langle u^\tau \rangle$ of the s95p equation of state is smaller than that for \textsc{neos} BQS and Duke at later times, but is slightly larger at earlier times before around $\tau=2$ fm. The orderings are consistent with those of the sound velocity and trace anomaly, considering that the average medium temperature is larger ($\sim 0.4$ GeV) at earlier times and smaller ($\sim 0.2$ GeV) at later times. 

The time evolution of the system's averaged momentum anisotropy
\begin{equation}
    \varepsilon_p = \frac{\sqrt{\langle T^{xx}-T^{yy} \rangle^2 + \langle 2T^{xy} \rangle^2}}{\langle T^{xx}+T^{yy} \rangle} ,
\end{equation}
in the same hydrodynamic setup is shown in Fig.~\ref{fig:comp_hydroevo_zeronB} (b). 
This quantity is closely related to the final elliptic momentum anisotropy of produced particles.
The differences in the momentum anisotropy between the equation of state models are rather small. The s95p result rises and falls slightly earlier than the others. The UH BQS equation of state has the largest momentum anisotropy at later times, followed by the BEST equation of state. \textsc{neos} BQS and Duke equations of state have similar $\varepsilon_p$, though the former is slightly larger than the latter around $\tau=4$ fm.

\begin{figure}[tb]
\centerline{
\includegraphics[width=2.4in,bb=0 0 576 432]{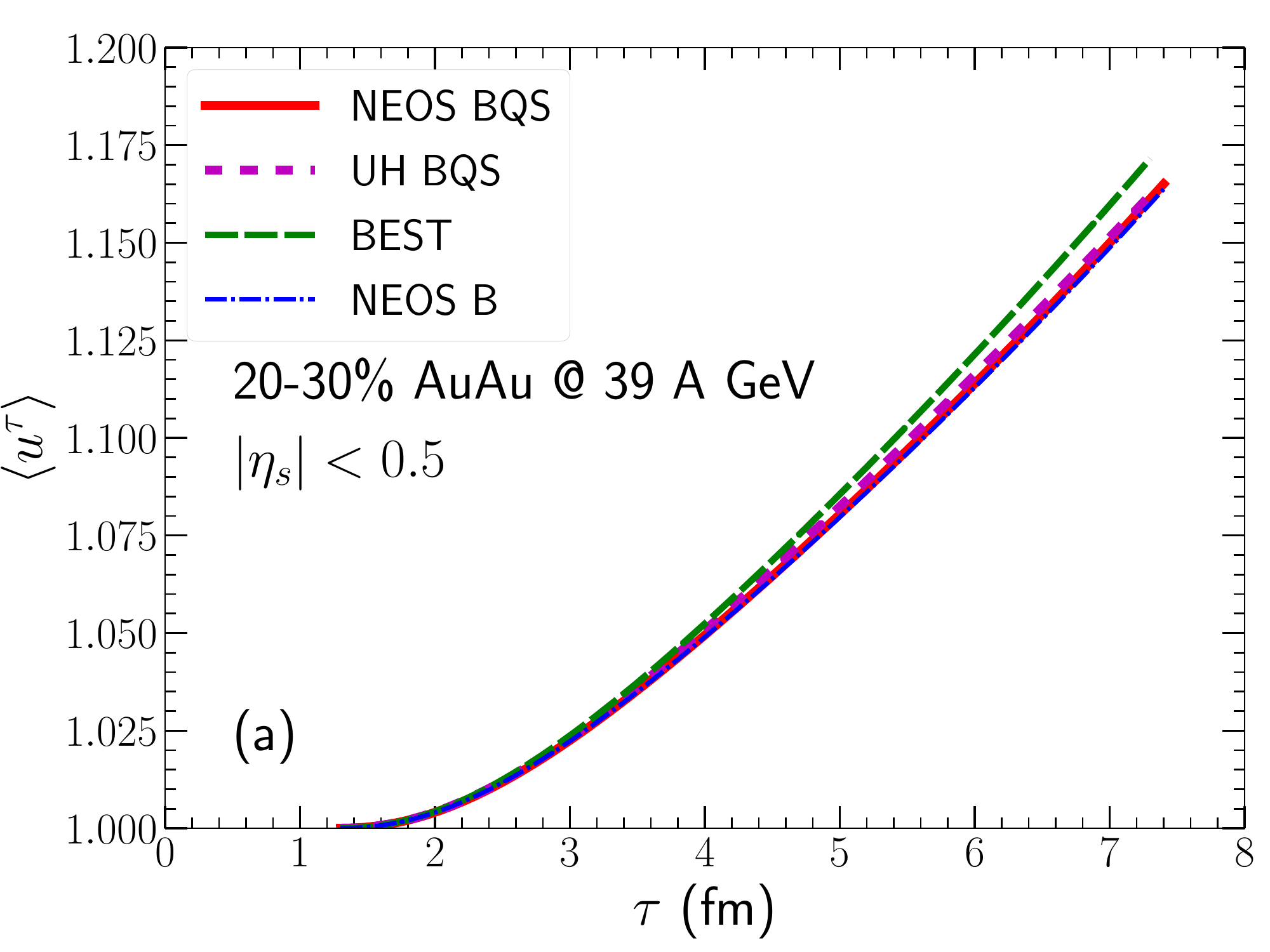}
\includegraphics[width=2.4in,bb=0 0 576 432]{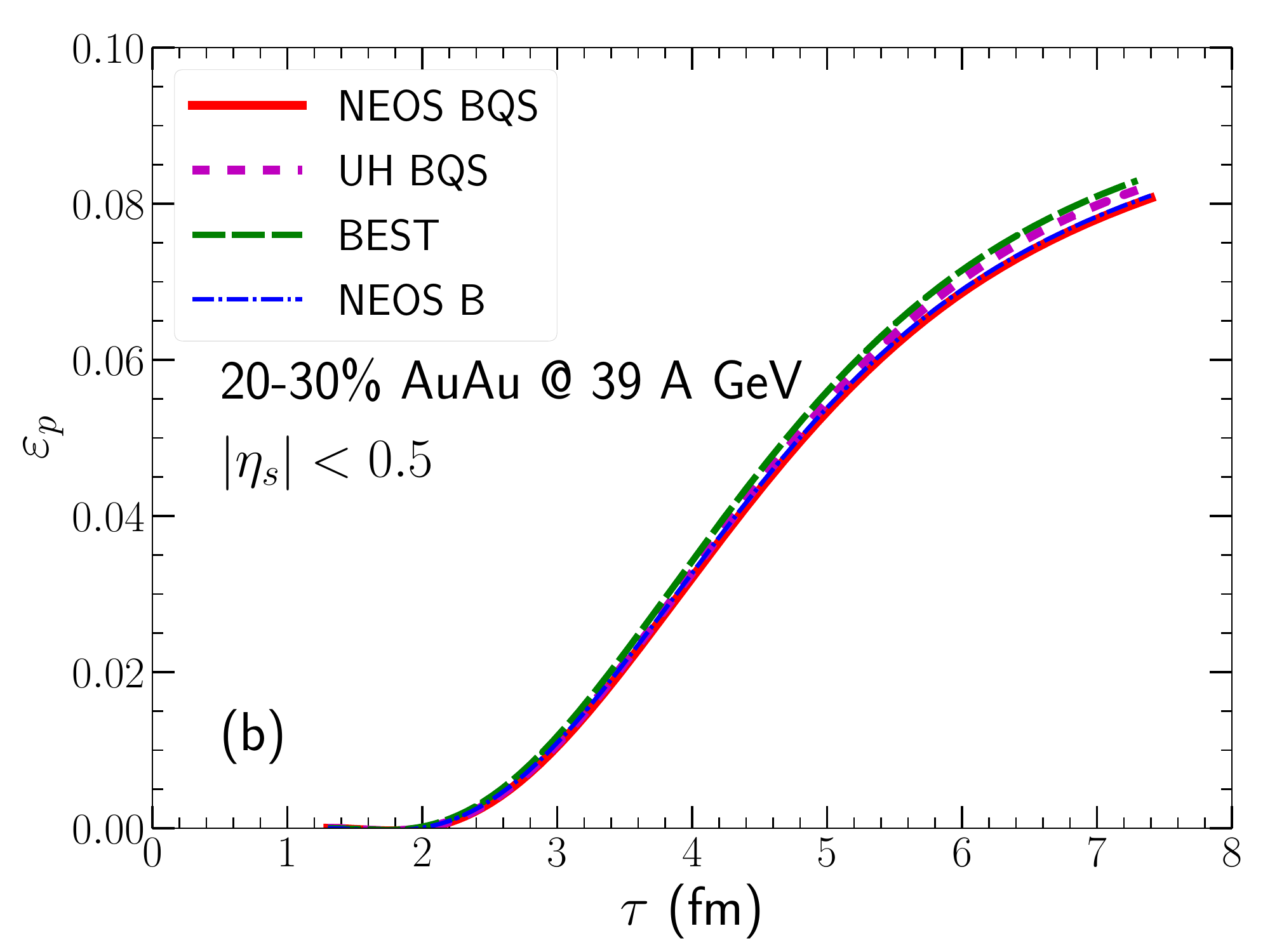}}
\caption{(a) The time evolution of averaged $u^\tau$ for an Au+Au collision in 20-30\% centrality at 39 GeV with different equations of state. (b) Similar comparison for the time evolution of the momentum anisotropy $\varepsilon_p$. \label{fig:comp_hydroevo_finiteB}}
\end{figure}

We make a similar comparison at finite net baryon density by simulating (3+1)D hydrodynamic evolution for 20-30\% Au+Au collisions at 39 GeV with the event-averaged initial condition \cite{Shen:2020jwv}. Figure~\ref{fig:comp_hydroevo_finiteB} shows that the four equations of state produce a very similar evolution for the development of hydrodynamic radial flow and the momentum anisotropy. Similar to the zero density case, the larger speed of sound in the BEST equation of state leads to slightly stronger radial flow compared to the \textsc{neos} and UH equations of state in Fig.~\ref{fig:comp_hydroevo_finiteB}(a). The system's momentum anisotropy at late time has the order BEST $>$ UH BQS $>$ \textsc{neos}.

\begin{figure}[tb]
\centerline{
\includegraphics[width=2.4in,bb=0 0 576 432]{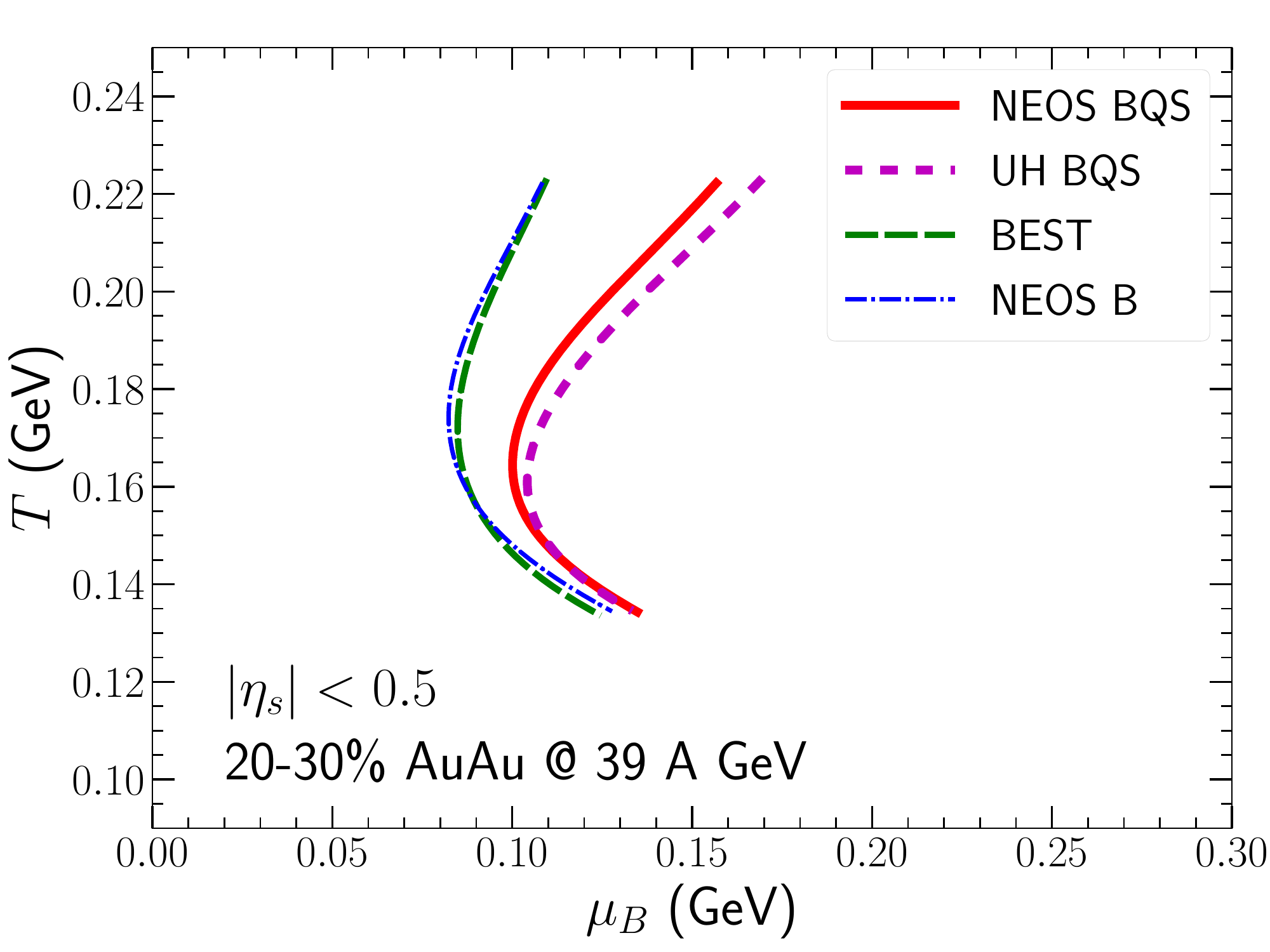}}
\caption{The averaged phase trajectories for a fireball at mid-rapidity in 20-30\% Au+Au collisions at 39 GeV. \label{fig:comp_phaseTrajectories}}
\end{figure}

Finally, we show the phase trajectories of mid-rapidity Au+Au collisions with the four equations of state in Figure~\ref{fig:comp_phaseTrajectories}. These trajectories are averaged over fluid cells from realistic (3+1)D hydrodynamic simulations. The difference among the four equations of state are in qualitative agreement with the difference in constant $s/n_B$ trajectories shown in Fig.~\ref{fig:comp_finitenB}. The strangeness neutrality condition moves the trajectories towards larger $\mu_B$ compared to those without this constraint. It indicates that, as mentioned earlier, having multiple conserved charges is phenomenologically important for the exploration of the QCD phase diagram, including the critical point search, as well as for the estimation of dissipative processes, such as baryon diffusion. The trajectory from the UH BQS has slightly larger $\mu_B$ values compared to the \textsc{neos} BQS in the QGP phase.

\section{Conclusion and summary}

We reviewed QCD equations of state at finite chemical potentials. All current models for the equation of state generally agree at zero densities, because of the advances in lattice QCD simulations, which all agree now that the quark-hadron transition is a crossover at around $T=155$-$160$ MeV, and the information from lattice QCD is used as input to determine parameters in the various phenomenological models. On the other hand, the finite-density structure of the QCD phase diagram, such as the critical point and first-order phase transition, is less well understood. Going beyond zero density is not directly possible on the lattice due to the fermion sign problem. Besides lattice based methods such as Taylor expansion or the use of imaginary chemical potentials, various approaches to obtain a finite density QCD equation of state have been proposed, including the perturbative QCD method, the Polyakov loop-extended Nambu-Jona-Lasinio model, and holographic conjecture. 

We have introduced the \textsc{neos} model where the three conserved charges in the strongly-interacting medium -- net baryon, electric charge and strangeness -- are explicitly considered. The model is built from a state-of-the-art lattice QCD equation of state and the second- and fourth-order susceptibilities from the lattice, together with the hadron resonance gas result, which includes all known hadrons and resonances with 
masses below 2 GeV. Lattice and hadron gas equations of state are connected near the quark-hadron transition in a thermodynamically consistent way to obtain a crossover-type equation of state at finite temperatures and chemical potentials. We have considered the strangeness neutrality condition $n_S=0$ and the electric charge-to-baryon ratio $n_Q=0.4n_B$, that reflect the situation in collisions of heavy nuclei to elucidate the effects of multiple conserved charges.

The multi-dimensional \textsc{neos} QCD equation of state has been included in the viscous hydrodynamic model of heavy-ion collisions \textsc{music} at intermediate relativistic energies. We showed in the comparison of the theoretical predictions with SPS experimental data of particle-antiparticle ratios that the model description is visibly improved when the strangeness neutrality condition is imposed for the hadrons with finite strangeness quantum numbers and -- through the interplay of conserved charges -- also for those with finite baryon number. The realistic electric charge-to-baryon ratio induces smaller effect because the electric charge chemical potential is small when heavy stable nuclei such as Au and Pb are used. Nevertheless, its inclusion leads to a correct description of the antipion-to-pion number ratio exceeding one in the observed data. Our results also clarify that in the beam energy scan one is really exploring the $T$-$\mu_B$-$\mu_Q$-$\mu_S$ phase diagram, instead of just one in the $T$-$\mu_B$ plane. This is important when extracting the information of the QCD medium properties and phase structures from experimental data.

We have then compared several models of the QCD equation of state used in relativistic hydrodynamic studies of nuclear collisions. In the zero density limit, the continuum limit results for the trace anomaly and sound velocity of the Wuppertal-Budapest collaboration and the HotQCD collaboration agree within their error bands. \textsc{neos} and Duke equations of state exhibit good agreement with the latter results while BEST and UH equations of state with the former results, as expected from their construction. The widely-used s95p-v1 has a larger trace anomaly and smaller sound velocity in the crossover region, and newer versions, such as s83s$_{18}$, should be used. The comparison of the constant $s/n_B$ trajectories in the phase diagram with different equations of state demonstrated that \textsc{neos} B and BEST equations of state are close to each other at finite density. Once the strangeness neutrality condition and the realistic charge-to-baryon ratio is taken into account, the trajectories are shifted to larger $\mu_B$. \text{neos} BQS and UH BQS equations of state behave similarly in the QGP phase with a slight difference coming likely from the choice of lattice data. The sound velocity differs between the two equations of state in the vicinity of the crossover, possibly because of the difference in the connection of the hadron resonance gas to the lattice QCD results. 

We performed several hydrodynamic simulations of heavy ion collisions using different equations of state and studied their effect on the time evolution of flow velocities and momentum anisotropies. At $\sqrt{s_{NN}} = 200$ GeV, the time evolution of the averaged time-like flow component $\langle u^\tau \rangle$ is in the order of BEST, UH BQS, \textsc{neos} BQS, Duke, and s95p, from the fastest to the slowest buildup (and largest to smallest final values), which is consistent with the ordering of the sound velocity in the zero density case near the crossover temperature. 
Comparing the averaged momentum anisotropies, final values for UH BQS and BEST are larger than for \textsc{neos} BQS and Duke, which in turn are larger than those for s95p. The differences in $\langle u^\tau \rangle$ at finite density at $\sqrt{s_{NN}} = 39$ GeV is rather small, but if closely observed, BEST produces the largest, followed by UH BQS and \textsc{neos} at later times because of the differences in the speed of sound. The average momentum anisotropy is also ordered similarly, but BEST and UH BQS are closest to each other. We also studied the trajectories of the average temperature and baryon chemical potential of the system, and found them to be qualitatively consistent with the constant $s/n_B$ trajectories of the corresponding equation of state models.

Progress in determining the nuclear equation of state at finite densities as been significant in the last several years. Advances in lattice QCD have allowed to move towards realistic modeling with finite chemical potentials and even studies including potential critical points are possible. Experimental advances have also been tremendous, both on the front of heavy ion collision beam energy scans and gravitational wave observations of neutron star (and black hole) binary systems, which will allow for ever improving constraints on the nuclear equation of state over a wide range in the phase diagram. 

A public version of the \textsc{neos} tabulated results is available online \cite{neos} for the use in relativistic hydrodynamic models and other related studies. Other codes/data are also publicly available for BEST \cite{best}, UH BQS \cite{uhbqs}, s95p-v1 \cite{s95p}, s83s$_{18}$ \cite{s83s}, and Duke \cite{Duke} equations of state.

\section*{Acknowledgments}

The authors thank Frithjof Karsch, Swagato Mukherjee, and Sayantan Sharma for useful discussions. AM is supported by JSPS KAKENHI Grant Number JP19K14722. BPS is supported under DOE Contract No. DE-SC0012704. CS is supported in part under DOE Contract No. DE-SC0013460 and in part by the National Science Foundation (NSF) under grant number PHY-2012922. This research used resources of the National Energy Research Scientific Computing Center, which is supported by the Office of Science of the U.S. Department of Energy under Contract No. DE-AC02-05CH11231. This work is supported in part by the U.S. Department of Energy, Office of Science, Office of Nuclear Physics, within the framework of the Beam Energy Scan Theory (BEST) Topical Collaboration.

\bibliography{neos}

\end{document}